\documentclass[prl,reprint,twocolumn,superscriptaddress,amsmath,amssymb,floatfix,longbibliography]{revtex4-2}
\usepackage{soul}
\usepackage{orcidlink}
\usepackage{graphicx}
\usepackage{epstopdf}
\usepackage[utf8]{inputenc}
\usepackage{amsmath,amsthm,amssymb}
\usepackage{latexsym}
\usepackage{bm}
\usepackage{dcolumn}
\usepackage{braket}
\usepackage[normalem]{ulem}
\usepackage{times}
\usepackage{hyperref}
\usepackage[section]{placeins}

\newcommand{\yms}{YbMnSb$_{2}$}
\newcommand{\ymb}{YbMnBi$_{2}$}

\newcommand{\cms}{Ca$_{0.6}$Sr$_{0.4}$MnSb$_{2}$}

\newcommand{\bR}{\mbox{\boldmath$R$}}
\newcommand{\bS}{\mbox{\boldmath$S$}}

\newcommand{\bq}{\mbox{\boldmath$q$}}
\newcommand{\bk}{\mbox{\boldmath$k$}}
\newcommand{\bki}{\mbox{\boldmath$k$}_{i}}
\newcommand{\bkf}{\mbox{\boldmath$k$}_{f}}
\newcommand{\bkone}{\mbox{\boldmath$k$}_{1}}
\newcommand{\bQ}{\mbox{\boldmath$Q$}}

\newcommand{\bM}{\mbox{\boldmath$M$}}

\newcommand{\btau}{\mbox{\boldmath$\tau$}}

\begin{document}

\title{A neutron diffraction study of spin wave excitations in \cms }
\title{Spin waves in Dirac semimetal \cms\ investigated by reinvented diffraction method}
\title{Spin waves in Dirac semimetal \cms\ investigated by neutron diffraction method}
\title{Spin waves in Dirac semimetal \cms\ investigated with neutrons by the diffraction method}

\author{Xiao Hu\orcidlink{0000-0002-0086-3127}} %
\affiliation{Condensed Matter Physics and Materials Science Division, Brookhaven National Laboratory, Upton, NY 11973, USA}

\author{Yan Wu\orcidlink{0000-0003-1566-614X}} %
\affiliation{Neutron Scattering Division, Oak Ridge National Laboratory, Oak Ridge, TN 37831, USA}

\author{Matthias D. Frontzek\orcidlink{0000-0001-8704-8928}} %
\affiliation{Neutron Scattering Division, Oak Ridge National Laboratory, Oak Ridge, TN 37831, USA}

\author{Zhixiang Hu\orcidlink{0000-0001-7925-2662}} %
\affiliation{Condensed Matter Physics and Materials Science Division, Brookhaven National Laboratory, Upton, NY 11973, USA}
\affiliation{Department of Material Science and Chemical Engineering, Stony Brook University, Stony Brook, NY, 11790, USA}

\author{Cedomir Petrovic\orcidlink{0000-0001-6063-1881}} %
\affiliation{Condensed Matter Physics and Materials Science Division, Brookhaven National Laboratory, Upton, NY 11973, USA}
\affiliation{Vin\v{c}a Institute of Nuclear Sciences, University of Belgrade, 11351 Vin\v{c}a, Belgrade, Serbia}

\author{John M. Tranquada\orcidlink{0000-0003-4984-8857}} %
\affiliation{Condensed Matter Physics and Materials Science Division, Brookhaven National Laboratory, Upton, NY 11973, USA}

\author{Igor A. Zaliznyak\orcidlink{0000-0002-8548-7924}} %
\email{zaliznyak@bnl.gov}
\affiliation{Condensed Matter Physics and Materials Science Division, Brookhaven National Laboratory, Upton, NY 11973, USA}

\begin{abstract}
{We report neutron diffraction measurements of \cms, a low-carrier-density Dirac semimetal in which the antiferromagnetic Mn layers are interleaved with Sb layers that host Dirac fermions. We have discovered that we can detect a good quality inelastic spin wave signal from a small (m~$\approx 0.28$~g) single crystal sample by the diffraction method, without energy analysis, using a neutron diffractometer with a position-sensitive area detector; the spin-waves appear as diffuse scattering that is shaped by energy-momentum conservation. By fitting this characteristic magnetic scattering to a spin-wave model, we refine all parameters of the model spin Hamiltonian, including the inter-plane interaction, through use of a three-dimensional measurement in reciprocal space. We also measure the temperature dependence of the spin waves, including the softening of the spin gap on approaching the N\'{e}el temperature, $T_N$. Not only do our results provide important new insights into an interplay of magnetism and Dirac electrons, they also establish a new, high-throughput approach to characterizing magnetic excitations on a modern diffractometer without direct energy analysis. Our work opens exciting new opportunities for the follow-up parametric and compositional studies on small, $\sim 0.1$~g crystals.
}
\end{abstract}

\date{\today}

\maketitle
\newpage

\noindent\emph{Introduction.}
Ever since its advent, neutron scattering has presented powerful tools for exploring the microscopic structure and dynamics of materials at the frontier of condensed matter research 
\cite{Brokhouse_RMP1995,Squires_book_1978,Boothroyd_book_2020,ZaliznyakLee_MNSChapter,Tranquada_2014,ZaliznyakTranquada_2014}. Recently, neutron studies have provided important insights into properties of Dirac and Weyl semimetals \cite{Armitage_2018}, which are intensely investigated due to their extremely high charge carrier mobilities, quantum transport at moderate temperatures, and topological and spin-polarized transport, which are all promising for technological applications \cite{Guo_PRB2014,Wang_2016,Zaliznyak_2017,Liu_2017,Soh_PRB2021,Rahn_PRB2017,Zhang_2019,Cai_PRB2020,Soh_PRB2019,Sapkota_2020,Hu_2023,Tobin_PRB2023}. Thanks to the neutron's specific sensitivity to magnetic interactions, neutron diffraction measurements have provided information on magnetic structures in Dirac semimetals hosting magnetic moments \cite{Guo_PRB2014,Wang_2016,Zaliznyak_2017,Liu_2017,Soh_PRB2021}, while spectroscopic inelastic neutron scattering (INS) has quantified the magnetic interactions in model spin Hamiltonians \cite{Rahn_PRB2017,Zhang_2019,Cai_PRB2020,Soh_PRB2019,Sapkota_2020,Hu_2023,Tobin_PRB2023}. Neutron diffraction measurements usually focus on elastic Bragg scattering and are performed on diffractometers without scattered neutron energy discrimination on small crystals, with typical masses
of $1 - 100$~mg. INS spectroscopic experiments, on the other hand, require scattered neutron energy discrimination in order to measure excitations in the four-dimensional $(\bQ, E)$ phase space (where $\hbar Q$ and $E$ are the momentum and energy transfers, respectively) and require large samples ($\gtrsim 1$~g) due to the two-orders-of-magnitude decrease in throughput that results from the scattered neutron energy analysis. This markedly hampers exploratory and composition-dependent INS studies compared to diffraction.

Among Dirac semimetals, 112 ternary pnictogens (A,R)MnX$_2$ (A = Ca, Sr, Ba; R = Yb, Eu; X = Bi, Sb) are particularly promising because they combine Dirac charge carriers with magnetism, allowing interaction between the two degrees of freedom that could lead to new interesting phenomena, such as Weyl states, tunable spin polarization, and the anomalous quantum Hall effect \cite{Guo_PRB2014,Wang_2016,Zaliznyak_2017,Liu_2017,Soh_PRB2021,Rahn_PRB2017,Zhang_2019,Cai_PRB2020,Soh_PRB2019,Sapkota_2020,Hu_2023,Tobin_PRB2023,Kefeng_2011,Ramankutty_2018,You_2019,Liu_PRB2019,Liu_2022,He_PRB_2017,Qiu_2018,Rong_PRB2021}. In these materials, the X layers hosting itinerant Dirac electrons are interleaved with strongly correlated, insulating, Mn-X layers which order antiferromagnetically near or above room temperature \cite{Guo_PRB2014,Wang_2016,Zaliznyak_2017,Liu_2017,Soh_PRB2021}. Both the inter-layer magnetic interactions and 3D charge transport require that Dirac charge carriers are coupled to Mn spins.
Intriguingly, neutron diffraction experiments \cite{Guo_PRB2014} found that character of the inter-layer spin coupling changes when the Ca in CaMnBi$_2$ is substituted with Sr. While antiferromagnetic Mn layers are stacked ferromagnetically in CaMnBi$_2$, the stacking is antiferromagnetic in SrMnBi$_2$. Follow-up INS spin wave measurements \cite{Rahn_PRB2017} established that Ca and Sr systems have very similar in-plane nearest- and next-nearest-neighbor couplings, $J_1$ and $J_2$, and an inter-layer coupling, $J_c$, of a nearly identical magnitude but of opposite sign, ferromagnetic ($J_c < 0$) in CaMnBi$_2$ and antiferromagnetic ($J_c > 0$) in SrMnBi$_2$. Whether this difference results from the change in the ionic size and electronegativity of the cations in-between the Mn-Bi layers, or a change of the crystal structure from $P4/nmm$ (CaMnBi$_2$) to $I4/mmm$ (SrMnBi$_2$), is unclear.

In order to explore the effect of alkali ion substitution on spin interactions and crystal and electronic structure, we carried out a neutron diffraction study on a small, $\approx 0.28$~g single crystal of Ca$_{0.6}$Sr$_{0.4}$MnSb$_2$, a member of the Sb-based 112 series similar to (Ca,Sr)MnBi$_2$ system but with a smaller spin-orbit coupling. The end member, SrMnSb$_2$, has been intensely studied since a putative ferromagnetism interacting with topological Dirac carriers was reported \cite{Liu_2017}, stirring substantial controversy \cite{Ramankutty_2018,You_2019,Liu_PRB2019}. It was resolved by ruling out an intrinsic ferromagnetic phase \cite{Liu_PRB2019}, which established SrMnSb$_2$ as an antiferromagnetic Dirac semimetal. SrMnSb$_2$ has an orthorhombic $Pnma$ lattice also found in YbMnSb$_2$, where it was shown that orthorhombicity has substantial effects on the electronic states near the Fermi level \cite{Bhoi_arXiv2023}. The studies of spin waves in SrMnSb$_2$ used INS on co-aligned muti-crystal arrays of m~$\gtrsim 0.5$~g \cite{Zhang_2019,Cai_PRB2020,Ning_PRB2024}.

While the initial goal of our neutron diffraction measurements was to investigate the effect of Sr/Ca substitution on magnetic and crystal structure of Ca$_{0.6}$Sr$_{0.4}$MnSb$_2$, the experiment also allowed us to obtain an accurate characterization of the spin waves to evaluate magnetic couplings in our small, diffraction-size Ca-doped crystal. This extraordinary development was enabled by recent advances in neutron instrumentation where modern diffractometers are equipped with large, highly pixellated position-sensitive detectors which simultaneously collect neutrons scattered within a large solid angle. Our results present a remarkable advancement of the ``diffraction method'', a neutron scattering technique for measuring magnetic excitations on a diffractometer with no energy analysis, which opens disruptive new directions for future studies.

The concept behind the diffraction method is not new -- in fact, it was used for the very first measurements of spin waves, which appear as bands of diffuse scattering near magnetic Bragg peaks \cite{Fishman_book_2018,ElliottLowde_1955,GoedkoopRiste_Nature1960}. However, on early day diffractometers with a single detector such measurements were severely limited. Only in the simplest cases, such as a ferromagnet where spin wave dispersion only depends on a single parameter, spin wave stiffness, can this interaction parameter be reliably measured \cite{Riste_1961,Frikkee_PhysicaB1966,Ferguson_PhysRev1967,Riste_1965,Shirane_PRL1965,Alperin_PhysRev1967}. Hence, following a brief period of popularity the method was abandoned in favor of spectroscopic measurements which, while limited by throughput and sample size, were providing more information. Here, we show how a quantitative change attained by increasing the number of detectors by several orders of magnitude has led to a qualitative advance that now makes the diffraction method quite powerful and an attractive complement to spectroscopy.

\begin{figure}[b!]
\vspace{-0.25in}
\includegraphics[width=1.0\columnwidth]{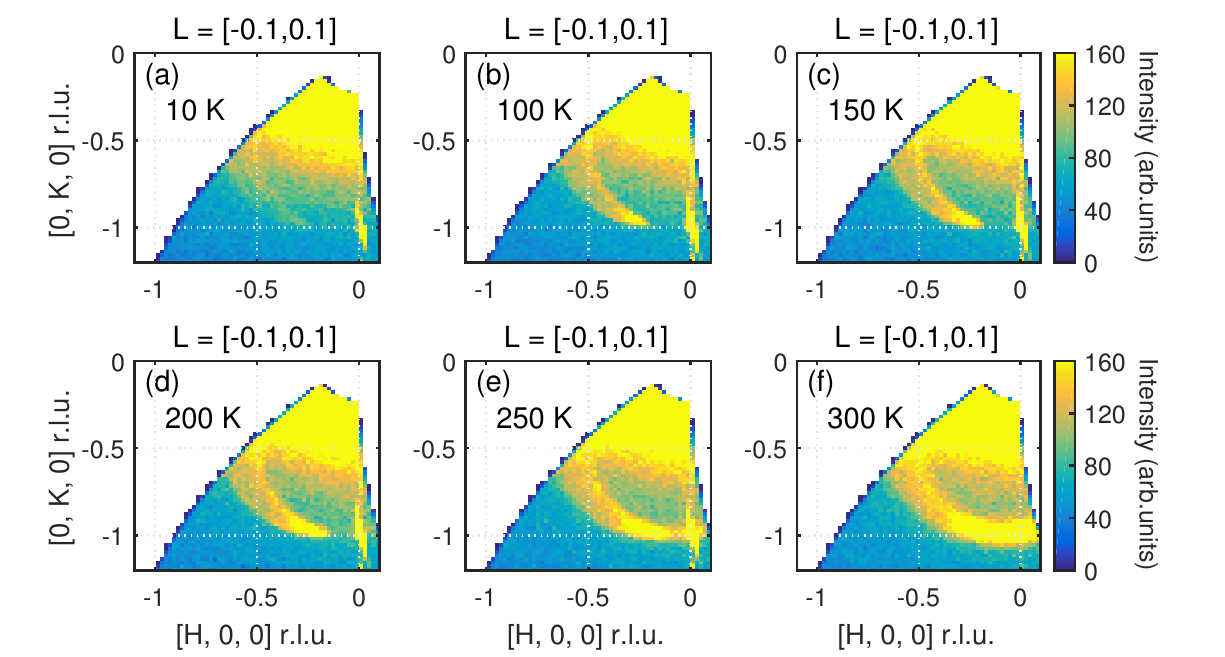}
\vspace{-0.1in}
\caption{{\bf Temperature-dependent neutron diffraction spectra of \cms.} (a--f) The neutron diffraction intensity measured with $\lambda = 1.486~\text\AA$ at 10, 100, 150, 200, 250, and 300 Kelvin averaged over $L\in[-0.1, 0.1]$; data bin size in $H$ and $K$ is $\pm 0.01$. Intense scattering increasing towards [H, K] = [0, 0] is the contamination from direct beam, which was fitted to a Gaussian and subtracted in analysis (see Fig.~\ref{Fig3:data&fit} and Supplementary Information \cite{Supplementary}). }
\label{Fig1:Tscan}
\end{figure}

\noindent\emph{Experimental Details.}
Single crystals of \cms\ were grown from Sb flux using the method described in {\cite{Wang_PRM2018}}.
Our neutron diffraction measurements were performed at the WAND$^2$ diffractometer at the High Flux Isotope Reactor, Oak Ridge National Laboratory, using incident neutron wavelength $\lambda = 1.486~\text\AA$ (energy $E_{i}=37$~meV). A single crystal of m~$\approx 0.28$~g was mounted in a closed cycle refrigerator and
aligned with the $(H, K, 0)$ reciprocal lattice zone in the horizontal scattering plane. The measurements were carried out by rotating the sample about the vertical axis in $0.1^{\circ}$ steps over a $180^{\circ}$ range.
Throughout the paper, we index the momentum transfer, $\bQ = (H, K, L)$, in reciprocal lattice units (r.l.u.) of the $P4/nmm$ lattice, $a =  b = 4.32(1)$~\AA, $c=10.85(1)$~\AA, referring to an undistorted YbMnSb$_{2}$ {\cite{Hu_2023,Tobin_PRB2023}}. The data reduction and binning to rectangular grid were performed using the MANTID package {\cite{arnold_mantid2014}} and the MDNorm algorithm {\cite{savici_MDnorm2022}}.

\begin{figure}[b!]
\vspace{-0.25in}
\includegraphics[width=1.0\columnwidth]{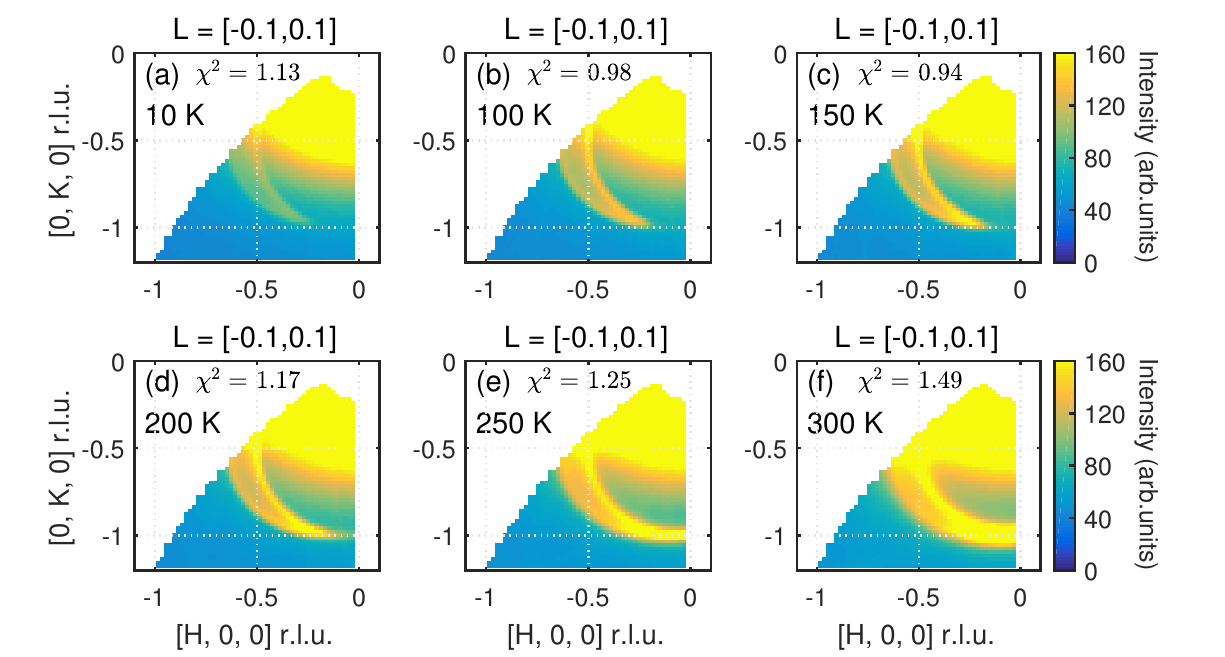}
\vspace{-0.1in}
\caption{{\bf  Fits of the temperature-dependent  neutron diffraction spectra of spin waves in \cms. }
(a--f) The simulated neutron diffraction intensity for 10, 100, 150, 200, 250, and 300 Kelvin corresponding to the data shown in Figure~\ref{Fig1:Tscan} obtained using the best fit parameters summarized in Fig.~\ref{Fig4:Tdependence} and Supplementary Table~S1. The direct beam background scattering modelled by a 2D Gaussian is included for direct comparison with Fig.~\ref{Fig1:Tscan}. }
\label{Fig2:Tscan_fit}
\end{figure}

\noindent\emph{Results and Analysis.}
Figure~\ref{Fig1:Tscan} presents high-resolution neutron diffraction spectra for \cms\ from 10 K to 300 K, close to its magnetic ordering transition, covering the first Brillouin zone at small diffraction angles where scattering is mainly magnetic. The data clearly reveal a distinctive band of diffuse scattering offset to the left of the antiferromagnetic (AFM) Bragg peak at $\bQ_{\rm AFM} = (0, -1, 0)$. Upon heating, the diffuse feature increases in intensity and gradually extends towards $\bQ_{\rm AFM}$, merging with it at 300~K. These characteristic behaviors immediately identify the crescent-shaped band as inelastic scattering from spin waves, which gains in intensity as the temperature increases. The offset from the magnetic Bragg peak indicates a gap in the spin wave spectrum, which closes on approaching the ordering temperature, $T_N$. The well-defined spin-waves with a gap $\Delta \approx 8 - 10$~meV at the AFM wave vector and consistent with the local-moment description and N\'eel antiferromagnetic order were observed by INS in the SrMnSb$_2$ end member \cite{Zhang_2019,Cai_PRB2020,Ning_PRB2024} and $A$MnBi$_{2}$ ($A$ = Ca, Sr) \cite{Rahn_PRB2017} sister materials.

%
%
To interpret our results, we consider neutron intensity measured in a diffraction experiment without scattered-neutron energy analysis. In such a case, each detector element collects all neutrons scattered by the sample in its direction independent of the scattered neutron energy, $E_f = \hbar^2 k_f^2/2m_n$ ($\bk_f$ is the scattered neutron wave vector, $m_n$ is neutron mass), so long as energy-momentum conservation laws are satisfied, $E = E_i - E_f = \hbar^2 ( k_{i}^2 - k_{f}^2)/{2m_n}$, $\bQ = \bki - \bkf$ ($E$ and $\bQ$ are energy and wave vector transferred to the sample). The corresponding intensity is given by,
\begin{equation}\label{IQ}
I(\bQ_{el}) = {A} \int_{-\infty}^{E_{i}}\frac{d^2 \sigma(\bQ, E)}{dE d\Omega} dE \, ,
\end{equation}
where $A$ is the normalization coefficient, ${d^2 \sigma(\bQ, E)}/{dE d\Omega}$ is the differential scattering cross-section of the sample, and the scattering direction is parameterized by the wave vector for elastic scattering, $\bQ_{el}$ (for details, see Supplementary Information \cite{Supplementary}). The integration in Eq.~\eqref{IQ} is along the trajectory, $\bQ = \bQ(E)$,
\begin{equation}\label{Q_Qel}
\bQ  = \bQ_{el}\sqrt{1-E/E_i} + \bk_i(1-\sqrt{1-E/E_i}) \,
\end{equation}
as determined by the conservation laws. For $E = 0$, $\bQ = \bQ_{el}$.

Equation~\eqref{IQ} shows that in a diffraction measurement without energy analysis the observed intensity is a projection of a four-dimensional $I(\bQ, E)$ onto a 3D $\bQ_{el}$-space along the energy integration trajectory defined by energy-momentum conservation, Eq.~\eqref{Q_Qel}. For each $\bQ_{el}$, the intensity measured by the corresponding detector element includes not only elastic scattering at $\bQ_{el}$, but also inelastic processes at different $\bQ$ that are present in the sample scattering cross-section. For example, spin-wave inelastic scattering at $\bQ_{\rm AFM}=(0,-1,0)$ and a spin-gap energy of $\Delta = 10$~meV would appear in our measurement at $\bQ_{el} \approx 1.17\bQ_{\rm AFM} - 0.17\bk_i$, offset from its true inelastic position, $\bQ = \bQ_{\rm AFM}$. The offset, $|\bQ_{el} - \bQ_{\rm AFM}|$, determines the spin-wave energy gap, $\Delta$, which is manifested as a $\bQ$-gap in the diffraction measurement. With the energy gap closing upon heating, $\Delta \rightarrow 0$, $\bQ_{el}\rightarrow \bQ_{\rm AFM}$ and the observed $\bQ$-gap gradually vanishes, as seen in Fig.~\ref{Fig1:Tscan}.

For the quantitative analysis of our data we use the spin-wave scattering cross-section of a layered Heisenberg antiferromagnet with the local-spin Hamiltonian, ${\cal H} = \sum_{ij} J_{ij} \bS_{i} \bS_{j} + D \sum_{i}(S^z_i)^2$, which was previously used for fitting spin waves in YbMnBi$_2$ and YbMnSb$_2$ \cite{Rahn_PRB2017,Soh_PRB2019,Cai_PRB2020,Sapkota_2020,Hu_2023}. Here, $J_{ij}$ includes exchange coupling between the nearest and next-nearest neighbors in the $ab$-plane, $J_1$ and $J_2$, and nearest-neighbors along the $c$-axis, $J_c$. $D < 0$ quantifies the uniaxial anisotropy for the Mn$^{2+}$ spins corresponding to an easy axis along the \textit{c} direction. Additionally, we use a damped-harmonic-oscillator (DHO) representation of the dynamical spin correlation function accounting for spin-wave damping via Lorentzian energy broadening parameter, $\gamma$ (see Refs. \onlinecite{Sapkota_2020,Hu_2023} for details). We thus fit the observed diffuse scattering intensity identified as spin waves to $I(\bQ_{el})$ obtained from Eq.~\eqref{IQ} using the spin-wave cross-section and corrected for the instrumental wave-vector resolution obtained by fitting the $(0,-1,0)$ Bragg peak to Gaussian profiles and bin size effects, which cause local averaging of the dispersion at each \bQ\ thus affecting the \bQ-distribution of spectral intensity.

In our analysis, we perform a global fit of three different $L$-slices of the data, averaged for $L \in [-0.6,-0.4]$, $[-0.35,-0.15]$, and $[-0.1,0.1]$, and with the binning size of $(\pm0.01, \pm0.01)$ in $(H,K)$, which provide adequate intensity for fitting (\emph{cf.} Fig.~\ref{Fig1:Tscan}) while minimizing the resolution effects (see Supplementary Information \cite{Supplementary} for details and representative fits). Figure~\ref{Fig2:Tscan_fit} presents the calculated spin wave intensity obtained from our model using the best fit parameters (Fig.~\ref{Fig4:Tdependence} and Table~S1 in \cite{Supplementary}) for the $L\in[-0.1, 0.1]$ slices of 10 K to 300 K data shown in Figure~\ref{Fig1:Tscan}. We observe that our model provides an excellent description of the data at all temperatures, with global fit chi-squared, $\chi^2 \sim 1$.
%
\begin{figure}[b!]
\vspace{-0.25in}
\includegraphics[width=1.0\columnwidth]{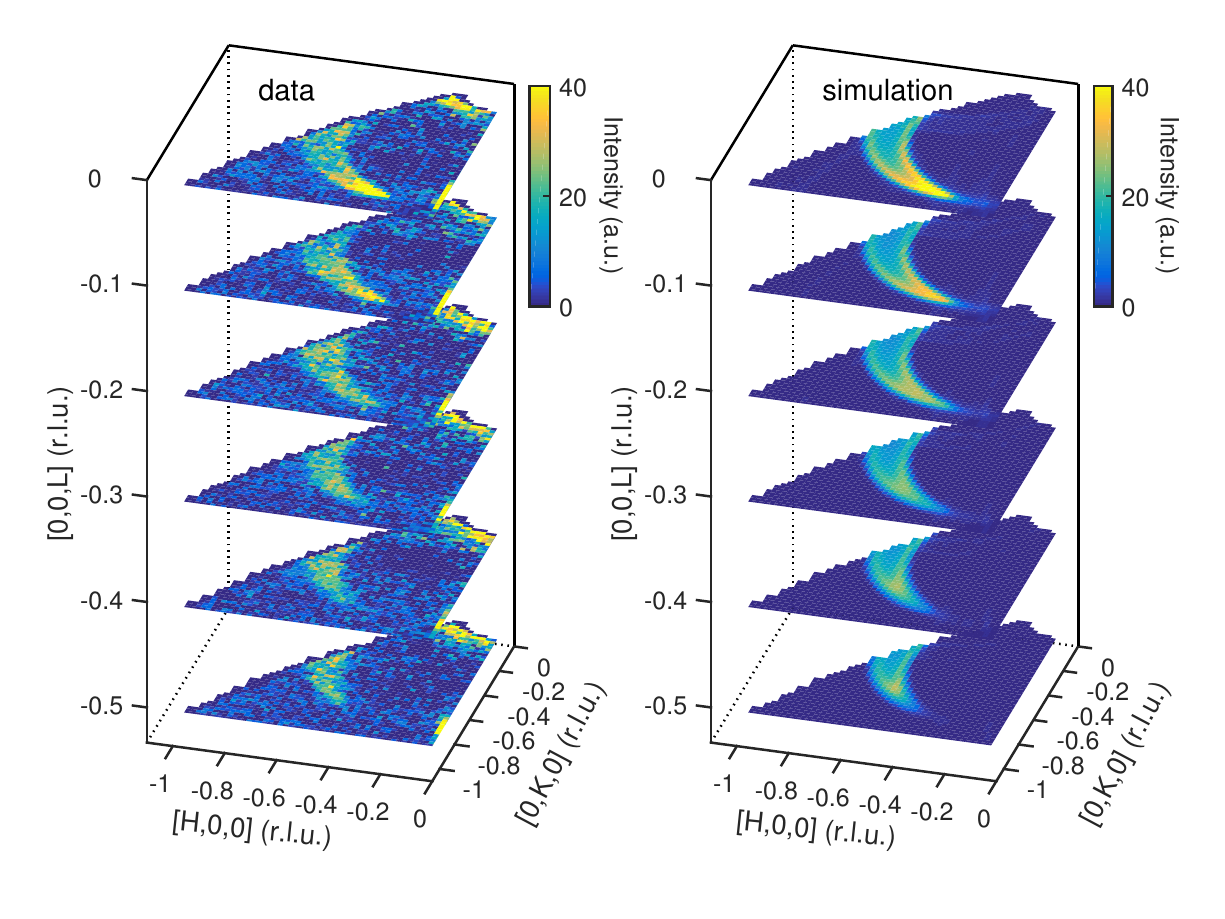}
\vspace{-0.25in}
\caption{{\bf $L-$dependence of spin waves in \cms\ at 100 K.} (Left) The spin wave spectra measured by neutron diffraction for different (vertical) inter-layer wave vector component, $L$, resolved along detector's vertical dimension. The data was averaged in the range $L = [0,-0.1,-0.2,-0.3,-0.4,-0.5] \pm 0.05$. The direct beam plus constant background intensity was evaluated by fitting to a 2D Gaussian profiles and subtracted from the data. (Right) The simulated spectra using fitted parameters of Model 2 in Table~\ref{Tab2ddisp}. }
\label{Fig3:data&fit}
\end{figure}

To understand the $L$-dependence and the damping effects {\cite{Sapkota_2020, Hu_2023}}, we focus on the 100 K data, which are still in the low-temperature regime with small thermal effects both on spin waves and the order parameter, but have stronger intensity compared to 10~K due to the detailed balance factor ($\approx 1/(1-1/e)\approx 1.5$ at $E \approx 10$~meV). We first fitted the data to Model 1 with the spectral broadening parameter fixed at a small value, $\gamma = 0.11$, essentially ignoring effects of spin wave damping, and then to Model 2 where the damping parameter, $\gamma$, was also varied. The interaction parameters obtained from the global fit of the three 2D constant-$L$ slices for both models agree well with each other and are summarized in Table~\ref{Tab2ddisp}. For Model 2 with optimized $\gamma = 0.9\pm0.1$ meV the fitted parameters have smaller uncertainties and the quality of the fit is slightly better, $\chi^{2}=0.98$ compared to $\chi^{2}=1.06$ for Model 1. The in-plane exchange couplings, $SJ_{1}=24.4\pm1.8$ meV, $SJ_{2}=8.2\pm1.2$ meV (Model 2), compare well with other ABX$_2$ systems {\cite{Zhang_2019,Cai_PRB2020,Rahn_PRB2017,Soh_PRB2019,Hu_2023,Sapkota_2020,Ning_PRB2024}}. 

Importantly, the inter-layer coupling obtained in our fits, $SJ_c \approx -0.17$~meV ($-0.19$~meV at 10~K, Supplementary Table~S1 \cite{Supplementary}), is ferromagnetic (FM) and similar in magnitude to that measured in the widely studied Sr end-member material, SrMnSb$_2$ \cite{Zhang_2019,Cai_PRB2020,Ning_PRB2024}. This answers the question posed at the outset of our study: there is little to no influence of Sr/Ca substitution in the interleaving cationic layer on the inter-layer coupling and magnetism in Ca$_{x}$Sr$_{1-x}$MnSb$_{2}$ family.

It should be noted that the values of $J_c$ refined in the previous spectroscopic neutron studies range from $-0.26$~meV \cite{Zhang_2019} to $-0.1$~meV \cite{Cai_PRB2020} to $-0.09$~meV \cite{Ning_PRB2024}, with similar scatter of the refined spin gap, $\Delta \approx$ 8.5, 6, and 10 meV, respectively. While these variations may partly arise from variations of stoichiometry other than Sr substitution, the other reason is the intensity-limited accuracy of the spectroscopic measurement. The diffraction method used in our work, on the other hand, is very sensitive to the spin gap and its dispersion across the vertical dimension of the high-resolution position-sensitive detector, which can be accurately refined even using a small, $\sim 0.1$~g sample, and in a much shorter measurement time. 

This is illustrated in Figure~\ref{Fig3:data&fit}, which presents our \cms\ diffraction data for different inter-layer wave vector transfer, $L$, together with the corresponding simulated intensity using the best fit parameters from Table~\ref{Tab2ddisp}. The spin wave $L$-dispersion and the energy gap translate into an offset of the measured diffraction intensity from the nominal magnetic Bragg peak position, $\bQ_{\rm AFM}$. Thanks to the detector's high wave vector resolution, this offset is clearly measurable, allowing [using Eq.~\eqref{Q_Qel}] to accurately read off the spin gap as a function of $L$ already from the raw data.

\begin{table}[t!]
\caption{\label{Tab2ddisp} Exchange coupling, uniaxial anisotropy, spin gap, and damping parameters for \cms\ at 100~K obtained from the global fit of two-dimensional data shown in Fig.~\ref{Fig1:Tscan} and Supplementary Figs.~S4 and S9 \cite{Supplementary} discussed in the text. Values in parentheses represent uncertainty with $95\%$ fitting confidence.}
\begin{ruledtabular}
\begin{tabular}{ccc}
& \mbox{Model 1} & \mbox{Model 2}\\
\hline
$SJ_1$ (meV)   & $23.8(36)$      & $24.4(18)$ \\
$SJ_2$ (meV)   & $7.5(25)$       & $8.2(12)$ \\
$SJ_c$ (meV)   & $-0.152(23)$    & $-0.166(14)$ \\
$SD$ (meV)     & $-0.068(10)$    & $-0.069(7)$ \\
$\Delta$ (meV) & $5.1(5)$        & $5.2(3)$ \\
$\gamma$ (meV) & $0.11$ (fixed)  & $0.9(1)$ \\

\end{tabular}
\end{ruledtabular}
\vspace{-0.25in}
\end{table}


The remarkable efficiency of the diffraction method allowed us to explore the temperature dependence of spin waves for temperatures up to $300$~K $\sim T_N$, all within a two-day experiment. The $T$-dependence of the $(0, -1, 0)$ magnetic Bragg peak, which corresponds to an in-plane AFM order of Mn spins with FM interplanar stacking (in agreement with the negative $J_c$ we refined), is shown in Figure~\ref{Fig4:Tdependence}(a). The order-parameter fit for $T > 250$~K reveals $T_N = 293(3)$~K and critical exponent, $\beta = 0.21(8)$, consistent with quasi-2D behavior and in agreement with similar results for SrMnSb$_2$ \cite{Zhang_2019}. The spin-wave parameters refined by fitting the diffraction intensity are presented in Figure~\ref{Fig4:Tdependence}(b--d) (see Figs.~\ref{Fig1:Tscan}, \ref{Fig2:Tscan_fit}, and Figures S2--S8 in \cite{Supplementary} for data and fits). On approaching the N\'{e}el temperature, the spin gap closes and we observe a critical increase of spin wave damping which, quite remarkably, can be confidently evaluated from our diffraction data [Fig.~\ref{Fig4:Tdependence}(b)]. The closure of the spin gap and disappearance of the inter-layer dispersion are reflected in the corresponding vanishing of the effective $T$-dependent inter-layer coupling and anisotropy parameters, $SJ_c(T)$ and $SD(T)$, shown in Fig~\ref{Fig4:Tdependence}(c). The in-plane spin interactions, on the other hand, remain $T$-independent [Fig.~\ref{Fig4:Tdependence}(d)], indicating the robustness of the 2D magnetism, expected for a layered quasi-2D system.
%
\begin{figure}[t!]
\includegraphics[width=1.0\columnwidth]{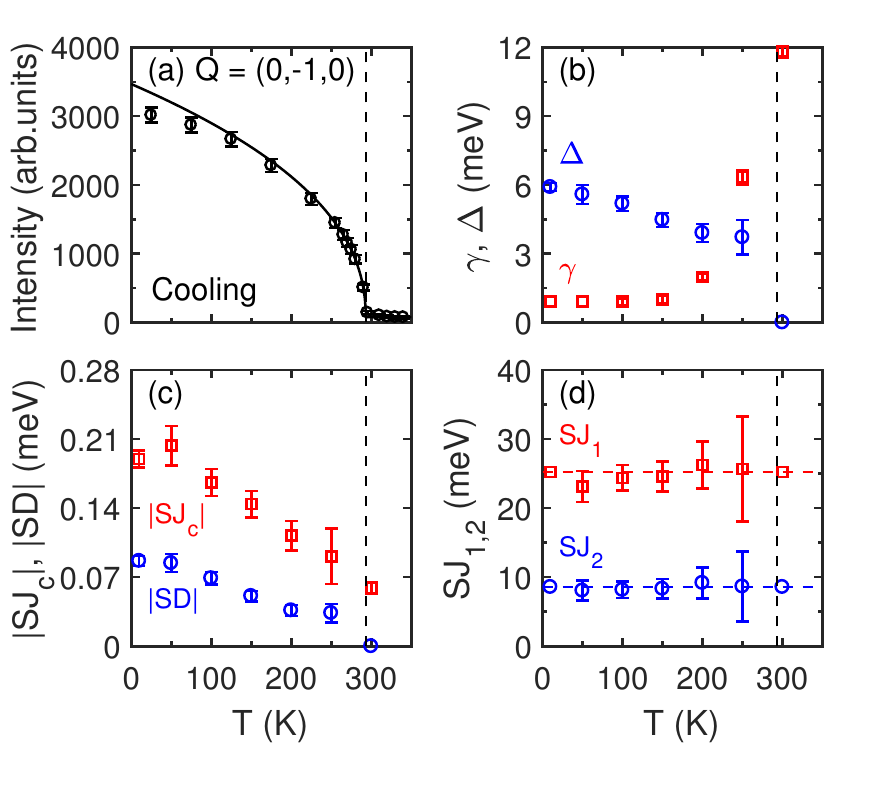}
\vspace{-0.25in}
\caption{{\bf Temperature dependence of magnetic order and spin interaction parameters in \cms.} (a) The intensity of $(0, -1, 0)$ magnetic Bragg peak; the solid line is fit to $I(T) = I_0 (1-T/T_N)^{2\beta}$ with $T_N = 293(3)$~K and $\beta = 0.21(8)$ quantifying the antiferromagnetic order parameter. (b) The vanishing spin gap, $\Delta$, and steeply increasing damping, $\gamma$, reveal critical behavior at $T \rightarrow T_N$. (c) The effective anisotropy parameter, $D(T)$, and the inter-layer coupling, $J_c(T)$, soften to zero together with the order parameter, indicating vanishing of the spin gap and the inter-layer dispersion in the disordered phase, at $T > T_N$. (d) The intra-layer couplings, $J_1$ and $J_2$, are $T$-independent, as expected for a quasi-2D magnetic system (values at 10~K and 300~K were fixed to the averages shown by horizontal broken lines).  The dashed vertical line in all panels marks $T_N$. The error bars mark one standard deviation. }
\label{Fig4:Tdependence}
\vspace{-0.25in}
\end{figure}

\noindent\emph{Summary and Conclusions.}
To summarize, our detailed investigation of spin waves in a layered Dirac semimetal, \cms, reveals that Sr/Ca substitution does not have measurable impact on the microscopic magnetism in the Ca$_{x}$Sr$_{1-x}$MnSb$_{2}$ family. This suggests that the switch from FM to AFM stacking observed in the sister Bi-based ABX$_2$ systems CaMnBi$_2$ and SrMnBi$_2$ \cite{Rahn_PRB2017} is not related to the change in the ionic size and electronegativity of the cations in-between the Mn-Bi layers, but likely results from the change of the crystal structure from $P4/nmm$ (CaMnBi$_2$) to $I4/mmm$ (SrMnBi$_2$) and the corresponding modification of the electronic structure. 

Our results were obtained by re-inventing the diffraction method \cite{ElliottLowde_1955,GoedkoopRiste_Nature1960,Riste_1961,Frikkee_PhysicaB1966,Ferguson_PhysRev1967,Riste_1965,Shirane_PRL1965,Alperin_PhysRev1967} using the recent advances in neutron diffractometer technology. Diffraction measurement provides a three-dimensional projection of the four-dimensional scattering function along the energy-momentum conservation trajectory, which is substantially more informative than a spectroscopic measurement of a 2D $(Q, E)$ projection from a polycrystalline sample that is often used in the absence of large single crystals. By abandoning the scattered neutron energy analysis, the diffraction method gains about a factor of $\sim 100$ in throughput compared to neutron spectroscopy while in many important cases still providing sufficient information for detailed refinement of microscopic models such as we report here. This advancement is significant, as it enables compositional and parametric studies of magnetic excitations on small, $\sim 100$~mg crystals that are not practical using conventional spectroscopy and thus opens exciting new avenues for materials research.

{\bf{Acknowledgements}} We would like to thank Dr. J. Fernandez-Baca for pointing out earlier references to the diffraction method. This work at Brookhaven National Laboratory was supported by Office of Basic Energy Sciences (BES), Division of Materials Sciences and Engineering,  U.S. Department of Energy (DOE), under contract DE-SC0012704. This research used resources at the High Flux Isotope Reactor, a DOE Office of Science User Facility operated by the Oak Ridge National Laboratory.

%



\begin{thebibliography}{42}%
\makeatletter
\providecommand \@ifxundefined [1]{%
 \@ifx{#1\undefined}
}%
\providecommand \@ifnum [1]{%
 \ifnum #1\expandafter \@firstoftwo
 \else \expandafter \@secondoftwo
 \fi
}%
\providecommand \@ifx [1]{%
 \ifx #1\expandafter \@firstoftwo
 \else \expandafter \@secondoftwo
 \fi
}%
\providecommand \natexlab [1]{#1}%
\providecommand \enquote  [1]{``#1''}%
\providecommand \bibnamefont  [1]{#1}%
\providecommand \bibfnamefont [1]{#1}%
\providecommand \citenamefont [1]{#1}%
\providecommand \href@noop [0]{\@secondoftwo}%
\providecommand \href [0]{\begingroup \@sanitize@url \@href}%
\providecommand \@href[1]{\@@startlink{#1}\@@href}%
\providecommand \@@href[1]{\endgroup#1\@@endlink}%
\providecommand \@sanitize@url [0]{\catcode `\\12\catcode `\$12\catcode
  `\&12\catcode `\#12\catcode `\^12\catcode `\_12\catcode `\%12\relax}%
\providecommand \@@startlink[1]{}%
\providecommand \@@endlink[0]{}%
\providecommand \url  [0]{\begingroup\@sanitize@url \@url }%
\providecommand \@url [1]{\endgroup\@href {#1}{\urlprefix }}%
\providecommand \urlprefix  [0]{URL }%
\providecommand \Eprint [0]{\href }%
\providecommand \doibase [0]{https://doi.org/}%
\providecommand \selectlanguage [0]{\@gobble}%
\providecommand \bibinfo  [0]{\@secondoftwo}%
\providecommand \bibfield  [0]{\@secondoftwo}%
\providecommand \translation [1]{[#1]}%
\providecommand \BibitemOpen [0]{}%
\providecommand \bibitemStop [0]{}%
\providecommand \bibitemNoStop [0]{.\EOS\space}%
\providecommand \EOS [0]{\spacefactor3000\relax}%
\providecommand \BibitemShut  [1]{\csname bibitem#1\endcsname}%
\let\auto@bib@innerbib\@empty
\bibitem [{\citenamefont {Brockhouse}(1995)}]{Brokhouse_RMP1995}%
  \BibitemOpen
  \bibfield  {author} {\bibinfo {author} {\bibfnamefont {B.~N.}\ \bibnamefont
  {Brockhouse}},\ }\bibfield  {title} {\bibinfo {title} {Slow neutron
  spectroscopy and the grand atlas of the physical world},\ }\href
  {https://doi.org/10.1103/RevModPhys.67.735} {\bibfield  {journal} {\bibinfo
  {journal} {Reviews of Modern Physics}\ }\textbf {\bibinfo {volume} {67}},\
  \bibinfo {pages} {735} (\bibinfo {year} {1995})}\BibitemShut {NoStop}%
\bibitem [{\citenamefont {Squires}(1978)}]{Squires_book_1978}%
  \BibitemOpen
  \bibfield  {author} {\bibinfo {author} {\bibfnamefont {G.~L.}\ \bibnamefont
  {Squires}},\ }\href {https://doi.org/10.1017/cbo9781139107808} {\emph
  {\bibinfo {title} {Introduction to the Theory of Thermal Neutron
  Scattering}}},\ \bibinfo {edition} {2012}\ ed.\ (\bibinfo  {publisher}
  {Cambridge University Press},\ \bibinfo {address} {England, UK},\ \bibinfo
  {year} {1978})\BibitemShut {NoStop}%
\bibitem [{\citenamefont {Boothroyd}(2020)}]{Boothroyd_book_2020}%
  \BibitemOpen
  \bibfield  {author} {\bibinfo {author} {\bibfnamefont {A.}~\bibnamefont
  {Boothroyd}},\ }\href {https://doi.org/10.1093/oso/9780198862314.001.0001}
  {\emph {\bibinfo {title} {Principles of Neutron Scattering from Condensed
  Matter}}}\ (\bibinfo  {publisher} {Oxford University Press},\ \bibinfo
  {address} {Oxford, UK},\ \bibinfo {year} {2020})\BibitemShut {NoStop}%
\bibitem [{\citenamefont {Zaliznyak}\ and\ \citenamefont
  {Lee}(2005)}]{ZaliznyakLee_MNSChapter}%
  \BibitemOpen
  \bibfield  {author} {\bibinfo {author} {\bibfnamefont {I.~A.}\ \bibnamefont
  {Zaliznyak}}\ and\ \bibinfo {author} {\bibfnamefont {S.-H.}\ \bibnamefont
  {Lee}},\ }\bibfield  {title} {\bibinfo {title} {Magnetic neutron
  scattering},\ }in\ \href {http://dx.doi.org/10.1007/0-387-23395-4_1} {\emph
  {\bibinfo {booktitle} {Modern Techniques for Characterizing Magnetic
  Materials}}},\ \bibinfo {editor} {edited by\ \bibinfo {editor} {\bibfnamefont
  {Y.}~\bibnamefont {Zhu}}}\ (\bibinfo  {publisher} {Springer US},\ \bibinfo
  {year} {2005})\ pp.\ \bibinfo {pages} {3--64}\BibitemShut {NoStop}%
\bibitem [{\citenamefont {Tranquada}\ \emph {et~al.}(2014)\citenamefont
  {Tranquada}, \citenamefont {Xu},\ and\ \citenamefont
  {Zaliznyak}}]{Tranquada_2014}%
  \BibitemOpen
  \bibfield  {author} {\bibinfo {author} {\bibfnamefont {J.~M.}\ \bibnamefont
  {Tranquada}}, \bibinfo {author} {\bibfnamefont {G.}~\bibnamefont {Xu}},\ and\
  \bibinfo {author} {\bibfnamefont {I.~A.}\ \bibnamefont {Zaliznyak}},\
  }\bibfield  {title} {\bibinfo {title} {Superconductivity, antiferromagnetism,
  and neutron scattering},\ }\href {https://doi.org/10.1016/j.jmmm.2013.09.029}
  {\bibfield  {journal} {\bibinfo  {journal} {Journal of Magnetism and Magnetic
  Materials}\ }\textbf {\bibinfo {volume} {350}},\ \bibinfo {pages} {148}
  (\bibinfo {year} {2014})}\BibitemShut {NoStop}%
\bibitem [{\citenamefont {Zaliznyak}\ and\ \citenamefont
  {Tranquada}(2014)}]{ZaliznyakTranquada_2014}%
  \BibitemOpen
  \bibfield  {author} {\bibinfo {author} {\bibfnamefont {I.~A.}\ \bibnamefont
  {Zaliznyak}}\ and\ \bibinfo {author} {\bibfnamefont {J.~M.}\ \bibnamefont
  {Tranquada}},\ }\bibfield  {title} {\bibinfo {title} {Neutron scattering and
  its application to strongly correlated systems},\ }in\ \href
  {https://doi.org/10.1007/978-3-662-44133-6_7} {\emph {\bibinfo {booktitle}
  {Springer Series in Solid-State Sciences}}}\ (\bibinfo  {publisher} {Springer
  Berlin Heidelberg},\ \bibinfo {year} {2014})\ pp.\ \bibinfo {pages}
  {205--235}\BibitemShut {NoStop}%
\bibitem [{\citenamefont {Armitage}\ \emph {et~al.}(2018)\citenamefont
  {Armitage}, \citenamefont {Mele},\ and\ \citenamefont
  {Vishwanath}}]{Armitage_2018}%
  \BibitemOpen
  \bibfield  {author} {\bibinfo {author} {\bibfnamefont {N.}~\bibnamefont
  {Armitage}}, \bibinfo {author} {\bibfnamefont {E.}~\bibnamefont {Mele}},\
  and\ \bibinfo {author} {\bibfnamefont {A.}~\bibnamefont {Vishwanath}},\
  }\bibfield  {title} {\bibinfo {title} {Weyl and Dirac semimetals in
  three-dimensional solids},\ }\href
  {https://doi.org/10.1103/revmodphys.90.015001} {\bibfield  {journal}
  {\bibinfo  {journal} {Reviews of Modern Physics}\ }\textbf {\bibinfo {volume}
  {90}},\ \bibinfo {pages} {015001} (\bibinfo {year} {2018})}\BibitemShut
  {NoStop}%
\bibitem [{\citenamefont {Guo}\ \emph {et~al.}(2014)\citenamefont {Guo},
  \citenamefont {Princep}, \citenamefont {Zhang}, \citenamefont {Manuel},
  \citenamefont {Khalyavin}, \citenamefont {Mazin}, \citenamefont {Shi},\ and\
  \citenamefont {Boothroyd}}]{Guo_PRB2014}%
  \BibitemOpen
  \bibfield  {author} {\bibinfo {author} {\bibfnamefont {Y.}~\bibnamefont
  {Guo}}, \bibinfo {author} {\bibfnamefont {A.}~\bibnamefont {Princep}},
  \bibinfo {author} {\bibfnamefont {X.}~\bibnamefont {Zhang}}, \bibinfo
  {author} {\bibfnamefont {P.}~\bibnamefont {Manuel}}, \bibinfo {author}
  {\bibfnamefont {D.}~\bibnamefont {Khalyavin}}, \bibinfo {author}
  {\bibfnamefont {I.}~\bibnamefont {Mazin}}, \bibinfo {author} {\bibfnamefont
  {Y.}~\bibnamefont {Shi}},\ and\ \bibinfo {author} {\bibfnamefont
  {A.}~\bibnamefont {Boothroyd}},\ }\bibfield  {title} {\bibinfo {title}
  {Coupling of magnetic order to planar Bi electrons in the anisotropic Dirac
  metals AMnBi$_2$ (A = Sr, Ca)},\ }\href
  {https://doi.org/10.1103/PhysRevB.90.075120} {\bibfield  {journal} {\bibinfo
  {journal} {Physical Review B}\ }\textbf {\bibinfo {volume} {90}},\ \bibinfo
  {pages} {075120} (\bibinfo {year} {2014})}\BibitemShut {NoStop}%
\bibitem [{\citenamefont {Wang}\ \emph {et~al.}(2016)\citenamefont {Wang},
  \citenamefont {Zaliznyak}, \citenamefont {Ren}, \citenamefont {Wu},
  \citenamefont {Graf}, \citenamefont {Garlea}, \citenamefont {Warren},
  \citenamefont {Bozin}, \citenamefont {Zhu},\ and\ \citenamefont
  {Petrovic}}]{Wang_2016}%
  \BibitemOpen
  \bibfield  {author} {\bibinfo {author} {\bibfnamefont {A.}~\bibnamefont
  {Wang}}, \bibinfo {author} {\bibfnamefont {I.}~\bibnamefont {Zaliznyak}},
  \bibinfo {author} {\bibfnamefont {W.}~\bibnamefont {Ren}}, \bibinfo {author}
  {\bibfnamefont {L.}~\bibnamefont {Wu}}, \bibinfo {author} {\bibfnamefont
  {D.}~\bibnamefont {Graf}}, \bibinfo {author} {\bibfnamefont {V.~O.}\
  \bibnamefont {Garlea}}, \bibinfo {author} {\bibfnamefont {J.~B.}\
  \bibnamefont {Warren}}, \bibinfo {author} {\bibfnamefont {E.}~\bibnamefont
  {Bozin}}, \bibinfo {author} {\bibfnamefont {Y.}~\bibnamefont {Zhu}},\ and\
  \bibinfo {author} {\bibfnamefont {C.}~\bibnamefont {Petrovic}},\ }\bibfield
  {title} {\bibinfo {title} {Magnetotransport study of Dirac fermions in
  YbMnBi$_{2}$ antiferromagnet},\ }\href
  {https://link.aps.org/doi/10.1103/PhysRevB.94.165161} {\bibfield  {journal}
  {\bibinfo  {journal} {Phys. Rev. B}\ }\textbf {\bibinfo {volume} {94}},\
  \bibinfo {pages} {165161} (\bibinfo {year} {2016})}\BibitemShut {NoStop}%
\bibitem [{\citenamefont {Zaliznyak}\ \emph {et~al.}(2017)\citenamefont
  {Zaliznyak}, \citenamefont {Savici}, \citenamefont {Garlea}, \citenamefont
  {Winn}, \citenamefont {Filges}, \citenamefont {Schneeloch}, \citenamefont
  {Tranquada}, \citenamefont {Gu}, \citenamefont {Wang},\ and\ \citenamefont
  {Petrovic}}]{Zaliznyak_2017}%
  \BibitemOpen
  \bibfield  {author} {\bibinfo {author} {\bibfnamefont {I.~A.}\ \bibnamefont
  {Zaliznyak}}, \bibinfo {author} {\bibfnamefont {A.~T.}\ \bibnamefont
  {Savici}}, \bibinfo {author} {\bibfnamefont {V.~O.}\ \bibnamefont {Garlea}},
  \bibinfo {author} {\bibfnamefont {B.}~\bibnamefont {Winn}}, \bibinfo {author}
  {\bibfnamefont {U.}~\bibnamefont {Filges}}, \bibinfo {author} {\bibfnamefont
  {J.}~\bibnamefont {Schneeloch}}, \bibinfo {author} {\bibfnamefont {J.~M.}\
  \bibnamefont {Tranquada}}, \bibinfo {author} {\bibfnamefont {G.}~\bibnamefont
  {Gu}}, \bibinfo {author} {\bibfnamefont {A.}~\bibnamefont {Wang}},\ and\
  \bibinfo {author} {\bibfnamefont {C.}~\bibnamefont {Petrovic}},\ }\bibfield
  {title} {\bibinfo {title} {Polarized neutron scattering on {HYSPEC}: the
  {HYbrid} {SPECtrometer} at {SNS}},\ }\href
  {https://doi.org/10.1088%2F1742-6596%2F862%2F1%2F012030} {\bibfield
  {journal} {\bibinfo  {journal} {Journal of Physics: Conference Series}\
  }\textbf {\bibinfo {volume} {862}},\ \bibinfo {pages} {012030} (\bibinfo
  {year} {2017})}\BibitemShut {NoStop}%
\bibitem [{\citenamefont {Liu}\ \emph {et~al.}(2017)\citenamefont {Liu},
  \citenamefont {Hu}, \citenamefont {Zhang}, \citenamefont {Graf},
  \citenamefont {Cao}, \citenamefont {Radmanesh}, \citenamefont {Adams},
  \citenamefont {Zhu}, \citenamefont {Cheng}, \citenamefont {Liu},
  \citenamefont {Phelan}, \citenamefont {Wei}, \citenamefont {Jaime},
  \citenamefont {Balakirev}, \citenamefont {Tennant}, \citenamefont {DiTusa},
  \citenamefont {Chiorescu}, \citenamefont {Spinu},\ and\ \citenamefont
  {Mao}}]{Liu_2017}%
  \BibitemOpen
  \bibfield  {author} {\bibinfo {author} {\bibfnamefont {J.~Y.}\ \bibnamefont
  {Liu}}, \bibinfo {author} {\bibfnamefont {J.}~\bibnamefont {Hu}}, \bibinfo
  {author} {\bibfnamefont {Q.}~\bibnamefont {Zhang}}, \bibinfo {author}
  {\bibfnamefont {D.}~\bibnamefont {Graf}}, \bibinfo {author} {\bibfnamefont
  {H.~B.}\ \bibnamefont {Cao}}, \bibinfo {author} {\bibfnamefont {S.~M.~A.}\
  \bibnamefont {Radmanesh}}, \bibinfo {author} {\bibfnamefont {D.~J.}\
  \bibnamefont {Adams}}, \bibinfo {author} {\bibfnamefont {Y.~L.}\ \bibnamefont
  {Zhu}}, \bibinfo {author} {\bibfnamefont {G.~F.}\ \bibnamefont {Cheng}},
  \bibinfo {author} {\bibfnamefont {X.}~\bibnamefont {Liu}}, \bibinfo {author}
  {\bibfnamefont {W.~A.}\ \bibnamefont {Phelan}}, \bibinfo {author}
  {\bibfnamefont {J.}~\bibnamefont {Wei}}, \bibinfo {author} {\bibfnamefont
  {M.}~\bibnamefont {Jaime}}, \bibinfo {author} {\bibfnamefont
  {F.}~\bibnamefont {Balakirev}}, \bibinfo {author} {\bibfnamefont {D.~A.}\
  \bibnamefont {Tennant}}, \bibinfo {author} {\bibfnamefont {J.~F.}\
  \bibnamefont {DiTusa}}, \bibinfo {author} {\bibfnamefont {I.}~\bibnamefont
  {Chiorescu}}, \bibinfo {author} {\bibfnamefont {L.}~\bibnamefont {Spinu}},\
  and\ \bibinfo {author} {\bibfnamefont {Z.~Q.}\ \bibnamefont {Mao}},\
  }\bibfield  {title} {\bibinfo {title} {A magnetic topological semimetal
  Sr$_{1-y}$Mn$_{1-z}$Sb$_{2}$ ($y, z < 0.1$)},\ }\href
  {https://doi.org/10.1038/nmat4953} {\bibfield  {journal} {\bibinfo  {journal}
  {Nature Materials}\ }\textbf {\bibinfo {volume} {16}},\ \bibinfo {pages}
  {905} (\bibinfo {year} {2017})}\BibitemShut {NoStop}%
\bibitem [{\citenamefont {Soh}\ \emph {et~al.}(2021)\citenamefont {Soh},
  \citenamefont {Tobin}, \citenamefont {Su}, \citenamefont {Zivkovic},
  \citenamefont {Ouladdiaf}, \citenamefont {Stunault}, \citenamefont
  {Rodríguez-Velamazán}, \citenamefont {Beauvois}, \citenamefont {Guo},\ and\
  \citenamefont {Boothroyd}}]{Soh_PRB2021}%
  \BibitemOpen
  \bibfield  {author} {\bibinfo {author} {\bibfnamefont {J.-R.}\ \bibnamefont
  {Soh}}, \bibinfo {author} {\bibfnamefont {S.~M.}\ \bibnamefont {Tobin}},
  \bibinfo {author} {\bibfnamefont {H.}~\bibnamefont {Su}}, \bibinfo {author}
  {\bibfnamefont {I.}~\bibnamefont {Zivkovic}}, \bibinfo {author}
  {\bibfnamefont {B.}~\bibnamefont {Ouladdiaf}}, \bibinfo {author}
  {\bibfnamefont {A.}~\bibnamefont {Stunault}}, \bibinfo {author}
  {\bibfnamefont {J.~A.}\ \bibnamefont {Rodríguez-Velamazán}}, \bibinfo
  {author} {\bibfnamefont {K.}~\bibnamefont {Beauvois}}, \bibinfo {author}
  {\bibfnamefont {Y.}~\bibnamefont {Guo}},\ and\ \bibinfo {author}
  {\bibfnamefont {A.~T.}\ \bibnamefont {Boothroyd}},\ }\bibfield  {title}
  {\bibinfo {title} {Magnetic structure of the topological semimetal
  YbMnSb$_{2}$},\ }\href {https://doi.org/10.1103/physrevb.104.l161103}
  {\bibfield  {journal} {\bibinfo  {journal} {Physical Review B}\ }\textbf
  {\bibinfo {volume} {104}},\ \bibinfo {pages} {l161103} (\bibinfo {year}
  {2021})}\BibitemShut {NoStop}%
\bibitem [{\citenamefont {Rahn}\ \emph {et~al.}(2017)\citenamefont {Rahn},
  \citenamefont {Princep}, \citenamefont {Piovano}, \citenamefont {Kulda},
  \citenamefont {Guo}, \citenamefont {Shi},\ and\ \citenamefont
  {Boothroyd}}]{Rahn_PRB2017}%
  \BibitemOpen
  \bibfield  {author} {\bibinfo {author} {\bibfnamefont {M.~C.}\ \bibnamefont
  {Rahn}}, \bibinfo {author} {\bibfnamefont {A.~J.}\ \bibnamefont {Princep}},
  \bibinfo {author} {\bibfnamefont {A.}~\bibnamefont {Piovano}}, \bibinfo
  {author} {\bibfnamefont {J.}~\bibnamefont {Kulda}}, \bibinfo {author}
  {\bibfnamefont {Y.~F.}\ \bibnamefont {Guo}}, \bibinfo {author} {\bibfnamefont
  {Y.~G.}\ \bibnamefont {Shi}},\ and\ \bibinfo {author} {\bibfnamefont {A.~T.}\
  \bibnamefont {Boothroyd}},\ }\bibfield  {title} {\bibinfo {title} {Spin
  dynamics in the antiferromagnetic phases of the Dirac metals
  $A{\mathrm{MnBi}}_{2}$ ($A=\mathrm{Sr}, \mathrm{Ca}$)},\ }\href
  {https://doi.org/10.1103/PhysRevB.95.134405} {\bibfield  {journal} {\bibinfo
  {journal} {Phys. Rev. B}\ }\textbf {\bibinfo {volume} {95}},\ \bibinfo
  {pages} {134405} (\bibinfo {year} {2017})}\BibitemShut {NoStop}%
\bibitem [{\citenamefont {Zhang}\ \emph {et~al.}(2019)\citenamefont {Zhang},
  \citenamefont {Okamoto}, \citenamefont {Stone}, \citenamefont {Liu},
  \citenamefont {Zhu}, \citenamefont {DiTusa}, \citenamefont {Mao},\ and\
  \citenamefont {Tennant}}]{Zhang_2019}%
  \BibitemOpen
  \bibfield  {author} {\bibinfo {author} {\bibfnamefont {Q.}~\bibnamefont
  {Zhang}}, \bibinfo {author} {\bibfnamefont {S.}~\bibnamefont {Okamoto}},
  \bibinfo {author} {\bibfnamefont {M.~B.}\ \bibnamefont {Stone}}, \bibinfo
  {author} {\bibfnamefont {J.}~\bibnamefont {Liu}}, \bibinfo {author}
  {\bibfnamefont {Y.}~\bibnamefont {Zhu}}, \bibinfo {author} {\bibfnamefont
  {J.}~\bibnamefont {DiTusa}}, \bibinfo {author} {\bibfnamefont
  {Z.}~\bibnamefont {Mao}},\ and\ \bibinfo {author} {\bibfnamefont {D.~A.}\
  \bibnamefont {Tennant}},\ }\bibfield  {title} {\bibinfo {title} {Influence of
  magnetism on Dirac semimetallic behavior in nonstoichiometric
  Sr$_{1-y}$Mn$_{1-z}$Sb$_{2}$ ($y \sim 0.07$, $z \sim 0.02$},\ }\href
  {https://doi.org/10.1103/physrevb.100.205105} {\bibfield  {journal} {\bibinfo
   {journal} {Physical Review B}\ }\textbf {\bibinfo {volume} {100}},\ \bibinfo
  {pages} {205105} (\bibinfo {year} {2019})}\BibitemShut {NoStop}%
\bibitem [{\citenamefont {Cai}\ \emph {et~al.}(2020)\citenamefont {Cai},
  \citenamefont {Bao}, \citenamefont {Wang}, \citenamefont {Ma}, \citenamefont
  {Dong}, \citenamefont {Shangguan}, \citenamefont {Wang}, \citenamefont {Ran},
  \citenamefont {Li}, \citenamefont {Kamazawa}, \citenamefont {Nakamura},
  \citenamefont {Adroja}, \citenamefont {Yu}, \citenamefont {Li},\ and\
  \citenamefont {Wen}}]{Cai_PRB2020}%
  \BibitemOpen
  \bibfield  {author} {\bibinfo {author} {\bibfnamefont {Z.}~\bibnamefont
  {Cai}}, \bibinfo {author} {\bibfnamefont {S.}~\bibnamefont {Bao}}, \bibinfo
  {author} {\bibfnamefont {W.}~\bibnamefont {Wang}}, \bibinfo {author}
  {\bibfnamefont {Z.}~\bibnamefont {Ma}}, \bibinfo {author} {\bibfnamefont
  {Z.-Y.}\ \bibnamefont {Dong}}, \bibinfo {author} {\bibfnamefont
  {Y.}~\bibnamefont {Shangguan}}, \bibinfo {author} {\bibfnamefont
  {J.}~\bibnamefont {Wang}}, \bibinfo {author} {\bibfnamefont {K.}~\bibnamefont
  {Ran}}, \bibinfo {author} {\bibfnamefont {S.}~\bibnamefont {Li}}, \bibinfo
  {author} {\bibfnamefont {K.}~\bibnamefont {Kamazawa}}, \bibinfo {author}
  {\bibfnamefont {M.}~\bibnamefont {Nakamura}}, \bibinfo {author}
  {\bibfnamefont {D.}~\bibnamefont {Adroja}}, \bibinfo {author} {\bibfnamefont
  {S.-L.}\ \bibnamefont {Yu}}, \bibinfo {author} {\bibfnamefont {J.-X.}\
  \bibnamefont {Li}},\ and\ \bibinfo {author} {\bibfnamefont {J.}~\bibnamefont
  {Wen}},\ }\bibfield  {title} {\bibinfo {title} {Spin dynamics of a magnetic
  Weyl semimetal Sr$_{1-x}$Mn$_{1-y}$Sb$_2$},\ }\href
  {https://doi.org/10.1103/PhysRevB.101.134408} {\bibfield  {journal} {\bibinfo
   {journal} {Phys. Rev. B}\ }\textbf {\bibinfo {volume} {101}},\ \bibinfo
  {pages} {134408} (\bibinfo {year} {2020})}\BibitemShut {NoStop}%
\bibitem [{\citenamefont {Soh}\ \emph {et~al.}(2019)\citenamefont {Soh},
  \citenamefont {Jacobsen}, \citenamefont {Ouladdiaf}, \citenamefont {Ivanov},
  \citenamefont {Piovano}, \citenamefont {Tejsner}, \citenamefont {Feng},
  \citenamefont {Wang}, \citenamefont {Su}, \citenamefont {Guo}, \citenamefont
  {Shi},\ and\ \citenamefont {Boothroyd}}]{Soh_PRB2019}%
  \BibitemOpen
  \bibfield  {author} {\bibinfo {author} {\bibfnamefont {J.-R.}\ \bibnamefont
  {Soh}}, \bibinfo {author} {\bibfnamefont {H.}~\bibnamefont {Jacobsen}},
  \bibinfo {author} {\bibfnamefont {B.}~\bibnamefont {Ouladdiaf}}, \bibinfo
  {author} {\bibfnamefont {A.}~\bibnamefont {Ivanov}}, \bibinfo {author}
  {\bibfnamefont {A.}~\bibnamefont {Piovano}}, \bibinfo {author} {\bibfnamefont
  {T.}~\bibnamefont {Tejsner}}, \bibinfo {author} {\bibfnamefont
  {Z.}~\bibnamefont {Feng}}, \bibinfo {author} {\bibfnamefont {H.}~\bibnamefont
  {Wang}}, \bibinfo {author} {\bibfnamefont {H.}~\bibnamefont {Su}}, \bibinfo
  {author} {\bibfnamefont {Y.}~\bibnamefont {Guo}}, \bibinfo {author}
  {\bibfnamefont {Y.}~\bibnamefont {Shi}},\ and\ \bibinfo {author}
  {\bibfnamefont {A.~T.}\ \bibnamefont {Boothroyd}},\ }\bibfield  {title}
  {\bibinfo {title} {Magnetic structure and excitations of the topological
  semimetal YbMnBi$_2$},\ }\href {https://doi.org/10.1103/PhysRevB.100.144431}
  {\bibfield  {journal} {\bibinfo  {journal} {Phys. Rev. B}\ }\textbf {\bibinfo
  {volume} {100}},\ \bibinfo {pages} {144431} (\bibinfo {year}
  {2019})}\BibitemShut {NoStop}%
\bibitem [{\citenamefont {Sapkota}\ \emph {et~al.}(2020)\citenamefont
  {Sapkota}, \citenamefont {Classen}, \citenamefont {Stone}, \citenamefont
  {Savici}, \citenamefont {Garlea}, \citenamefont {Wang}, \citenamefont
  {Tranquada}, \citenamefont {Petrovic},\ and\ \citenamefont
  {Zaliznyak}}]{Sapkota_2020}%
  \BibitemOpen
  \bibfield  {author} {\bibinfo {author} {\bibfnamefont {A.}~\bibnamefont
  {Sapkota}}, \bibinfo {author} {\bibfnamefont {L.}~\bibnamefont {Classen}},
  \bibinfo {author} {\bibfnamefont {M.~B.}\ \bibnamefont {Stone}}, \bibinfo
  {author} {\bibfnamefont {A.~T.}\ \bibnamefont {Savici}}, \bibinfo {author}
  {\bibfnamefont {V.~O.}\ \bibnamefont {Garlea}}, \bibinfo {author}
  {\bibfnamefont {A.}~\bibnamefont {Wang}}, \bibinfo {author} {\bibfnamefont
  {J.~M.}\ \bibnamefont {Tranquada}}, \bibinfo {author} {\bibfnamefont
  {C.}~\bibnamefont {Petrovic}},\ and\ \bibinfo {author} {\bibfnamefont
  {I.~A.}\ \bibnamefont {Zaliznyak}},\ }\bibfield  {title} {\bibinfo {title}
  {Signatures of coupling between spin waves and Dirac fermions in
  YbMnBi$_2$},\ }\href {https://doi.org/10.1103/PhysRevB.101.041111} {\bibfield
   {journal} {\bibinfo  {journal} {Phys. Rev. B}\ }\textbf {\bibinfo {volume}
  {101}},\ \bibinfo {pages} {041111} (\bibinfo {year} {2020})}\BibitemShut
  {NoStop}%
\bibitem [{\citenamefont {Hu}\ \emph {et~al.}(2023)\citenamefont {Hu},
  \citenamefont {Sapkota}, \citenamefont {Hu}, \citenamefont {Savici},
  \citenamefont {Kolesnikov}, \citenamefont {Tranquada}, \citenamefont
  {Petrovic},\ and\ \citenamefont {Zaliznyak}}]{Hu_2023}%
  \BibitemOpen
  \bibfield  {author} {\bibinfo {author} {\bibfnamefont {X.}~\bibnamefont
  {Hu}}, \bibinfo {author} {\bibfnamefont {A.}~\bibnamefont {Sapkota}},
  \bibinfo {author} {\bibfnamefont {Z.}~\bibnamefont {Hu}}, \bibinfo {author}
  {\bibfnamefont {A.~T.}\ \bibnamefont {Savici}}, \bibinfo {author}
  {\bibfnamefont {A.~I.}\ \bibnamefont {Kolesnikov}}, \bibinfo {author}
  {\bibfnamefont {J.~M.}\ \bibnamefont {Tranquada}}, \bibinfo {author}
  {\bibfnamefont {C.}~\bibnamefont {Petrovic}},\ and\ \bibinfo {author}
  {\bibfnamefont {I.~A.}\ \bibnamefont {Zaliznyak}},\ }\bibfield  {title}
  {\bibinfo {title} {Coupling of magnetism and Dirac fermions in YbMnSb$_2$},\
  }\href {https://doi.org/10.1103/PhysRevB.107.L201117} {\bibfield  {journal}
  {\bibinfo  {journal} {Phys. Rev. B}\ }\textbf {\bibinfo {volume} {107}},\
  \bibinfo {pages} {L201117} (\bibinfo {year} {2023})}\BibitemShut {NoStop}%
\bibitem [{\citenamefont {Tobin}\ \emph {et~al.}(2023)\citenamefont {Tobin},
  \citenamefont {Soh}, \citenamefont {Su}, \citenamefont {Piovano},
  \citenamefont {Stunault}, \citenamefont {Rodríguez-Velamazán},
  \citenamefont {Guo},\ and\ \citenamefont {Boothroyd}}]{Tobin_PRB2023}%
  \BibitemOpen
  \bibfield  {author} {\bibinfo {author} {\bibfnamefont {S.~M.}\ \bibnamefont
  {Tobin}}, \bibinfo {author} {\bibfnamefont {J.-R.}\ \bibnamefont {Soh}},
  \bibinfo {author} {\bibfnamefont {H.}~\bibnamefont {Su}}, \bibinfo {author}
  {\bibfnamefont {A.}~\bibnamefont {Piovano}}, \bibinfo {author} {\bibfnamefont
  {A.}~\bibnamefont {Stunault}}, \bibinfo {author} {\bibfnamefont {J.~A.}\
  \bibnamefont {Rodríguez-Velamazán}}, \bibinfo {author} {\bibfnamefont
  {Y.}~\bibnamefont {Guo}},\ and\ \bibinfo {author} {\bibfnamefont {A.~T.}\
  \bibnamefont {Boothroyd}},\ }\bibfield  {title} {\bibinfo {title} {Magnetic
  excitations in the topological semimetal YbMnSb$_{2}$},\ }\href
  {https://doi.org/10.1103/physrevb.107.195146} {\bibfield  {journal} {\bibinfo
   {journal} {Physical Review B}\ }\textbf {\bibinfo {volume} {107}},\ \bibinfo
  {pages} {195146} (\bibinfo {year} {2023})}\BibitemShut {NoStop}%
\bibitem [{\citenamefont {Wang}\ \emph {et~al.}(2011)\citenamefont {Wang},
  \citenamefont {Graf}, \citenamefont {Lei}, \citenamefont {Tozer},\ and\
  \citenamefont {Petrovic}}]{Kefeng_2011}%
  \BibitemOpen
  \bibfield  {author} {\bibinfo {author} {\bibfnamefont {K.}~\bibnamefont
  {Wang}}, \bibinfo {author} {\bibfnamefont {D.}~\bibnamefont {Graf}}, \bibinfo
  {author} {\bibfnamefont {H.}~\bibnamefont {Lei}}, \bibinfo {author}
  {\bibfnamefont {S.~W.}\ \bibnamefont {Tozer}},\ and\ \bibinfo {author}
  {\bibfnamefont {C.}~\bibnamefont {Petrovic}},\ }\bibfield  {title} {\bibinfo
  {title} {Quantum transport of two-dimensional Dirac fermions in
  SrMnBi${}_{2}$},\ }\href
  {https://link.aps.org/doi/10.1103/PhysRevB.84.220401} {\bibfield  {journal}
  {\bibinfo  {journal} {Phys. Rev. B}\ }\textbf {\bibinfo {volume} {84}},\
  \bibinfo {pages} {220401} (\bibinfo {year} {2011})}\BibitemShut {NoStop}%
\bibitem [{\citenamefont {Ramankutty}\ \emph {et~al.}(2018)\citenamefont
  {Ramankutty}, \citenamefont {Henke}, \citenamefont {Schiphorst},
  \citenamefont {Nutakki}, \citenamefont {Bron}, \citenamefont
  {Araizi-Kanoutas}, \citenamefont {Mishra}, \citenamefont {Li}, \citenamefont
  {Huang}, \citenamefont {Kim}, \citenamefont {Hoesch}, \citenamefont
  {Schlueter}, \citenamefont {Lee}, \citenamefont {de~Visser}, \citenamefont
  {Zhong}, \citenamefont {van Wezel}, \citenamefont {van Heumen},\ and\
  \citenamefont {Golden}}]{Ramankutty_2018}%
  \BibitemOpen
  \bibfield  {author} {\bibinfo {author} {\bibfnamefont {S.~V.}\ \bibnamefont
  {Ramankutty}}, \bibinfo {author} {\bibfnamefont {J.}~\bibnamefont {Henke}},
  \bibinfo {author} {\bibfnamefont {A.}~\bibnamefont {Schiphorst}}, \bibinfo
  {author} {\bibfnamefont {R.}~\bibnamefont {Nutakki}}, \bibinfo {author}
  {\bibfnamefont {S.}~\bibnamefont {Bron}}, \bibinfo {author} {\bibfnamefont
  {G.}~\bibnamefont {Araizi-Kanoutas}}, \bibinfo {author} {\bibfnamefont
  {S.}~\bibnamefont {Mishra}}, \bibinfo {author} {\bibfnamefont
  {L.}~\bibnamefont {Li}}, \bibinfo {author} {\bibfnamefont {Y.}~\bibnamefont
  {Huang}}, \bibinfo {author} {\bibfnamefont {T.}~\bibnamefont {Kim}}, \bibinfo
  {author} {\bibfnamefont {M.}~\bibnamefont {Hoesch}}, \bibinfo {author}
  {\bibfnamefont {C.}~\bibnamefont {Schlueter}}, \bibinfo {author}
  {\bibfnamefont {T.-L.}\ \bibnamefont {Lee}}, \bibinfo {author} {\bibfnamefont
  {A.}~\bibnamefont {de~Visser}}, \bibinfo {author} {\bibfnamefont
  {Z.}~\bibnamefont {Zhong}}, \bibinfo {author} {\bibfnamefont
  {J.}~\bibnamefont {van Wezel}}, \bibinfo {author} {\bibfnamefont
  {E.}~\bibnamefont {van Heumen}},\ and\ \bibinfo {author} {\bibfnamefont
  {M.}~\bibnamefont {Golden}},\ }\bibfield  {title} {\bibinfo {title}
  {Electronic structure of the candidate 2D Dirac semimetal SrMnSb$_{2}$: a
  combined experimental and theoretical study},\ }\bibfield  {journal}
  {\bibinfo  {journal} {SciPost Physics}\ }\textbf {\bibinfo {volume} {4}},\
  \href {https://doi.org/10.21468/scipostphys.4.2.010}
  {10.21468/scipostphys.4.2.010} (\bibinfo {year} {2018})\BibitemShut {NoStop}%
\bibitem [{\citenamefont {You}\ \emph {et~al.}(2019)\citenamefont {You},
  \citenamefont {Lee}, \citenamefont {Choi}, \citenamefont {Jo}, \citenamefont
  {Shim},\ and\ \citenamefont {Kim}}]{You_2019}%
  \BibitemOpen
  \bibfield  {author} {\bibinfo {author} {\bibfnamefont {J.~S.}\ \bibnamefont
  {You}}, \bibinfo {author} {\bibfnamefont {I.}~\bibnamefont {Lee}}, \bibinfo
  {author} {\bibfnamefont {E.~S.}\ \bibnamefont {Choi}}, \bibinfo {author}
  {\bibfnamefont {Y.~J.}\ \bibnamefont {Jo}}, \bibinfo {author} {\bibfnamefont
  {J.~H.}\ \bibnamefont {Shim}},\ and\ \bibinfo {author} {\bibfnamefont
  {J.~S.}\ \bibnamefont {Kim}},\ }\bibfield  {title} {\bibinfo {title}
  {Shubnikov-de Haas oscillations of massive Dirac fermions in a Dirac
  antiferromagnet SrMnSb$_2$},\ }\href
  {https://doi.org/10.1016/j.cap.2018.10.022} {\bibfield  {journal} {\bibinfo
  {journal} {Current Applied Physics}\ }\textbf {\bibinfo {volume} {19}},\
  \bibinfo {pages} {230} (\bibinfo {year} {2019})}\BibitemShut {NoStop}%
\bibitem [{\citenamefont {Liu}\ \emph {et~al.}(2019)\citenamefont {Liu},
  \citenamefont {Islam}, \citenamefont {Dennis}, \citenamefont {Tian},
  \citenamefont {Ueland}, \citenamefont {McQueeney},\ and\ \citenamefont
  {Vaknin}}]{Liu_PRB2019}%
  \BibitemOpen
  \bibfield  {author} {\bibinfo {author} {\bibfnamefont {Y.}~\bibnamefont
  {Liu}}, \bibinfo {author} {\bibfnamefont {F.}~\bibnamefont {Islam}}, \bibinfo
  {author} {\bibfnamefont {K.~W.}\ \bibnamefont {Dennis}}, \bibinfo {author}
  {\bibfnamefont {W.}~\bibnamefont {Tian}}, \bibinfo {author} {\bibfnamefont
  {B.~G.}\ \bibnamefont {Ueland}}, \bibinfo {author} {\bibfnamefont {R.~J.}\
  \bibnamefont {McQueeney}},\ and\ \bibinfo {author} {\bibfnamefont
  {D.}~\bibnamefont {Vaknin}},\ }\bibfield  {title} {\bibinfo {title} {Hole
  doping and antiferromagnetic correlations above the Néel temperature of the
  topological semimetal (Sr$_{1-x}$K$_x$)MnSb$_2$},\ }\href
  {https://doi.org/10.1103/physrevb.100.014437} {\bibfield  {journal} {\bibinfo
   {journal} {Physical Review B}\ }\textbf {\bibinfo {volume} {100}},\ \bibinfo
  {pages} {014437} (\bibinfo {year} {2019})}\BibitemShut {NoStop}%
\bibitem [{\citenamefont {Liu}\ \emph {et~al.}(2022)\citenamefont {Liu},
  \citenamefont {Fu}, \citenamefont {Cheng}, \citenamefont {Zhu}, \citenamefont
  {He}, \citenamefont {Liu}, \citenamefont {Li},\ and\ \citenamefont
  {Luo}}]{Liu_2022}%
  \BibitemOpen
  \bibfield  {author} {\bibinfo {author} {\bibfnamefont {B.}~\bibnamefont
  {Liu}}, \bibinfo {author} {\bibfnamefont {Y.}~\bibnamefont {Fu}}, \bibinfo
  {author} {\bibfnamefont {J.}~\bibnamefont {Cheng}}, \bibinfo {author}
  {\bibfnamefont {W.}~\bibnamefont {Zhu}}, \bibinfo {author} {\bibfnamefont
  {J.}~\bibnamefont {He}}, \bibinfo {author} {\bibfnamefont {C.}~\bibnamefont
  {Liu}}, \bibinfo {author} {\bibfnamefont {L.}~\bibnamefont {Li}},\ and\
  \bibinfo {author} {\bibfnamefont {Y.}~\bibnamefont {Luo}},\ }\bibfield
  {title} {\bibinfo {title} {Physical properties of antiferromagnetic Dirac
  semimetal SrMnSb$_{2}$},\ }\href {https://doi.org/10.1007/s10948-022-06403-5}
  {\bibfield  {journal} {\bibinfo  {journal} {Journal of Superconductivity and
  Novel Magnetism}\ }\textbf {\bibinfo {volume} {35}},\ \bibinfo {pages} {3263}
  (\bibinfo {year} {2022})}\BibitemShut {NoStop}%
\bibitem [{\citenamefont {He}\ \emph {et~al.}(2017)\citenamefont {He},
  \citenamefont {Fu}, \citenamefont {Zhao}, \citenamefont {Liang},
  \citenamefont {Chen}, \citenamefont {Leng}, \citenamefont {Wang},
  \citenamefont {Li}, \citenamefont {Zhang}, \citenamefont {Xue}, \citenamefont
  {Li}, \citenamefont {Zhang}, \citenamefont {Ren},\ and\ \citenamefont
  {Chen}}]{He_PRB_2017}%
  \BibitemOpen
  \bibfield  {author} {\bibinfo {author} {\bibfnamefont {J.~B.}\ \bibnamefont
  {He}}, \bibinfo {author} {\bibfnamefont {Y.}~\bibnamefont {Fu}}, \bibinfo
  {author} {\bibfnamefont {L.~X.}\ \bibnamefont {Zhao}}, \bibinfo {author}
  {\bibfnamefont {H.}~\bibnamefont {Liang}}, \bibinfo {author} {\bibfnamefont
  {D.}~\bibnamefont {Chen}}, \bibinfo {author} {\bibfnamefont {Y.~M.}\
  \bibnamefont {Leng}}, \bibinfo {author} {\bibfnamefont {X.~M.}\ \bibnamefont
  {Wang}}, \bibinfo {author} {\bibfnamefont {J.}~\bibnamefont {Li}}, \bibinfo
  {author} {\bibfnamefont {S.}~\bibnamefont {Zhang}}, \bibinfo {author}
  {\bibfnamefont {M.~Q.}\ \bibnamefont {Xue}}, \bibinfo {author} {\bibfnamefont
  {C.~H.}\ \bibnamefont {Li}}, \bibinfo {author} {\bibfnamefont
  {P.}~\bibnamefont {Zhang}}, \bibinfo {author} {\bibfnamefont {Z.~A.}\
  \bibnamefont {Ren}},\ and\ \bibinfo {author} {\bibfnamefont {G.~F.}\
  \bibnamefont {Chen}},\ }\bibfield  {title} {\bibinfo {title}
  {Quasi-two-dimensional massless Dirac fermions in CaMnSb$_2$},\ }\href
  {https://doi.org/10.1103/PhysRevB.95.045128} {\bibfield  {journal} {\bibinfo
  {journal} {Phys. Rev. B}\ }\textbf {\bibinfo {volume} {95}},\ \bibinfo
  {pages} {045128} (\bibinfo {year} {2017})}\BibitemShut {NoStop}%
\bibitem [{\citenamefont {Qiu}\ \emph {et~al.}(2018)\citenamefont {Qiu},
  \citenamefont {Le}, \citenamefont {Dai}, \citenamefont {Xu}, \citenamefont
  {He}, \citenamefont {Yang}, \citenamefont {Chen}, \citenamefont {Hu},\ and\
  \citenamefont {Qiu}}]{Qiu_2018}%
  \BibitemOpen
  \bibfield  {author} {\bibinfo {author} {\bibfnamefont {Z.}~\bibnamefont
  {Qiu}}, \bibinfo {author} {\bibfnamefont {C.}~\bibnamefont {Le}}, \bibinfo
  {author} {\bibfnamefont {Y.}~\bibnamefont {Dai}}, \bibinfo {author}
  {\bibfnamefont {B.}~\bibnamefont {Xu}}, \bibinfo {author} {\bibfnamefont
  {J.~B.}\ \bibnamefont {He}}, \bibinfo {author} {\bibfnamefont
  {R.}~\bibnamefont {Yang}}, \bibinfo {author} {\bibfnamefont {G.}~\bibnamefont
  {Chen}}, \bibinfo {author} {\bibfnamefont {J.}~\bibnamefont {Hu}},\ and\
  \bibinfo {author} {\bibfnamefont {X.}~\bibnamefont {Qiu}},\ }\bibfield
  {title} {\bibinfo {title} {Infrared spectroscopic studies of the topological
  properties in CaMnSb$_{2}$},\ }\href
  {https://doi.org/10.1103/physrevb.98.115151} {\bibfield  {journal} {\bibinfo
  {journal} {Physical Review B}\ }\textbf {\bibinfo {volume} {98}},\ \bibinfo
  {pages} {115151} (\bibinfo {year} {2018})}\BibitemShut {NoStop}%
\bibitem [{\citenamefont {Rong}\ \emph {et~al.}(2021)\citenamefont {Rong},
  \citenamefont {Zhou}, \citenamefont {He}, \citenamefont {Song}, \citenamefont
  {Huang}, \citenamefont {Hu}, \citenamefont {Xu}, \citenamefont {Cai},
  \citenamefont {Chen}, \citenamefont {Li}, \citenamefont {Wang}, \citenamefont
  {Zhao}, \citenamefont {Zhu}, \citenamefont {Liu}, \citenamefont {Xu},
  \citenamefont {Chen}, \citenamefont {Weng},\ and\ \citenamefont
  {Zhou}}]{Rong_PRB2021}%
  \BibitemOpen
  \bibfield  {author} {\bibinfo {author} {\bibfnamefont {H.}~\bibnamefont
  {Rong}}, \bibinfo {author} {\bibfnamefont {L.}~\bibnamefont {Zhou}}, \bibinfo
  {author} {\bibfnamefont {J.}~\bibnamefont {He}}, \bibinfo {author}
  {\bibfnamefont {C.}~\bibnamefont {Song}}, \bibinfo {author} {\bibfnamefont
  {J.}~\bibnamefont {Huang}}, \bibinfo {author} {\bibfnamefont
  {C.}~\bibnamefont {Hu}}, \bibinfo {author} {\bibfnamefont {Y.}~\bibnamefont
  {Xu}}, \bibinfo {author} {\bibfnamefont {Y.}~\bibnamefont {Cai}}, \bibinfo
  {author} {\bibfnamefont {H.}~\bibnamefont {Chen}}, \bibinfo {author}
  {\bibfnamefont {C.}~\bibnamefont {Li}}, \bibinfo {author} {\bibfnamefont
  {Q.}~\bibnamefont {Wang}}, \bibinfo {author} {\bibfnamefont {L.}~\bibnamefont
  {Zhao}}, \bibinfo {author} {\bibfnamefont {Z.}~\bibnamefont {Zhu}}, \bibinfo
  {author} {\bibfnamefont {G.}~\bibnamefont {Liu}}, \bibinfo {author}
  {\bibfnamefont {Z.}~\bibnamefont {Xu}}, \bibinfo {author} {\bibfnamefont
  {G.}~\bibnamefont {Chen}}, \bibinfo {author} {\bibfnamefont {H.}~\bibnamefont
  {Weng}},\ and\ \bibinfo {author} {\bibfnamefont {X.~J.}\ \bibnamefont
  {Zhou}},\ }\bibfield  {title} {\bibinfo {title} {Electronic structure
  examination of the topological properties of CaMnSb$_2$ by angle-resolved
  photoemission spectroscopy},\ }\href
  {https://doi.org/10.1103/PhysRevB.103.245104} {\bibfield  {journal} {\bibinfo
   {journal} {Phys. Rev. B}\ }\textbf {\bibinfo {volume} {103}},\ \bibinfo
  {pages} {245104} (\bibinfo {year} {2021})}\BibitemShut {NoStop}%
\bibitem [{\citenamefont {Bhoi}\ \emph {et~al.}(2023)\citenamefont {Bhoi},
  \citenamefont {Ye}, \citenamefont {Ma}, \citenamefont {Shen}, \citenamefont
  {Maurya}, \citenamefont {Kasamatsu}, \citenamefont {Misawa}, \citenamefont
  {Yoshimi}, \citenamefont {Nakajima}, \citenamefont {Matsuda},\ and\
  \citenamefont {Uwatoko}}]{Bhoi_arXiv2023}%
  \BibitemOpen
  \bibfield  {author} {\bibinfo {author} {\bibfnamefont {D.}~\bibnamefont
  {Bhoi}}, \bibinfo {author} {\bibfnamefont {F.}~\bibnamefont {Ye}}, \bibinfo
  {author} {\bibfnamefont {H.}~\bibnamefont {Ma}}, \bibinfo {author}
  {\bibfnamefont {X.}~\bibnamefont {Shen}}, \bibinfo {author} {\bibfnamefont
  {A.}~\bibnamefont {Maurya}}, \bibinfo {author} {\bibfnamefont
  {S.}~\bibnamefont {Kasamatsu}}, \bibinfo {author} {\bibfnamefont
  {T.}~\bibnamefont {Misawa}}, \bibinfo {author} {\bibfnamefont
  {K.}~\bibnamefont {Yoshimi}}, \bibinfo {author} {\bibfnamefont
  {T.}~\bibnamefont {Nakajima}}, \bibinfo {author} {\bibfnamefont
  {M.}~\bibnamefont {Matsuda}},\ and\ \bibinfo {author} {\bibfnamefont
  {Y.}~\bibnamefont {Uwatoko}},\ }\bibfield  {title} {\bibinfo {title} {Fermi
  surface reconstruction due to the orthorhombic distortion in Dirac semimetal
  YbMnSb$_2$},\ }\href {https://doi.org/10.48550/arXiv.2306.12732} {\bibfield
  {journal} {\bibinfo  {journal} {arXiv e-prints}\ ,\ \bibinfo {pages}
  {arXiv:2306.12732}} (\bibinfo {year} {2023})}\BibitemShut {NoStop}%
\bibitem [{\citenamefont {Ning}\ \emph {et~al.}(2024)\citenamefont {Ning},
  \citenamefont {Li}, \citenamefont {Tang}, \citenamefont {Banerjee},
  \citenamefont {Fanelli}, \citenamefont {Abernathy}, \citenamefont {Liu},
  \citenamefont {Ueland}, \citenamefont {McQueeney},\ and\ \citenamefont
  {Ke}}]{Ning_PRB2024}%
  \BibitemOpen
  \bibfield  {author} {\bibinfo {author} {\bibfnamefont {Z.}~\bibnamefont
  {Ning}}, \bibinfo {author} {\bibfnamefont {B.}~\bibnamefont {Li}}, \bibinfo
  {author} {\bibfnamefont {W.}~\bibnamefont {Tang}}, \bibinfo {author}
  {\bibfnamefont {A.}~\bibnamefont {Banerjee}}, \bibinfo {author}
  {\bibfnamefont {V.}~\bibnamefont {Fanelli}}, \bibinfo {author} {\bibfnamefont
  {D.~L.}\ \bibnamefont {Abernathy}}, \bibinfo {author} {\bibfnamefont
  {Y.}~\bibnamefont {Liu}}, \bibinfo {author} {\bibfnamefont {B.~G.}\
  \bibnamefont {Ueland}}, \bibinfo {author} {\bibfnamefont {R.~J.}\
  \bibnamefont {McQueeney}},\ and\ \bibinfo {author} {\bibfnamefont
  {L.}~\bibnamefont {Ke}},\ }\bibfield  {title} {\bibinfo {title} {Magnetic
  interactions and excitations in SrMnSb$_2$},\ }\href
  {https://doi.org/10.1103/physrevb.109.214414} {\bibfield  {journal} {\bibinfo
   {journal} {Physical Review B}\ }\textbf {\bibinfo {volume} {109}},\ \bibinfo
  {pages} {214414} (\bibinfo {year} {2024})}\BibitemShut {NoStop}%
\bibitem [{\citenamefont {Fishman}\ \emph {et~al.}(2018)\citenamefont
  {Fishman}, \citenamefont {Fernandez-Baca},\ and\ \citenamefont
  {R{\~o}{\~o}m}}]{Fishman_book_2018}%
  \BibitemOpen
  \bibfield  {author} {\bibinfo {author} {\bibfnamefont {R.~S.}\ \bibnamefont
  {Fishman}}, \bibinfo {author} {\bibfnamefont {J.~A.}\ \bibnamefont
  {Fernandez-Baca}},\ and\ \bibinfo {author} {\bibfnamefont {T.}~\bibnamefont
  {R{\~o}{\~o}m}},\ }\href {https://doi.org/10.1088/978-1-64327-114-9} {\emph
  {\bibinfo {title} {Spin-Wave Theory and its Applications to Neutron
  Scattering and THz Spectroscopy}}}\ (\bibinfo  {publisher} {Morgan {\&}
  Claypool Publishers},\ \bibinfo {address} {USA},\ \bibinfo {year}
  {2018})\BibitemShut {NoStop}%
\bibitem [{\citenamefont {Elliott}\ and\ \citenamefont
  {Lowde}(1955)}]{ElliottLowde_1955}%
  \BibitemOpen
  \bibfield  {author} {\bibinfo {author} {\bibfnamefont {R.~J.}\ \bibnamefont
  {Elliott}}\ and\ \bibinfo {author} {\bibfnamefont {R.~D.}\ \bibnamefont
  {Lowde}},\ }\bibfield  {title} {\bibinfo {title} {The inelastic scattering of
  neutrons by magnetic spin waves},\ }\href
  {https://doi.org/10.1098/rspa.1955.0112} {\bibfield  {journal} {\bibinfo
  {journal} {Proceedings of the Royal Society of London. Series A. Mathematical
  and Physical Sciences}\ }\textbf {\bibinfo {volume} {230}},\ \bibinfo {pages}
  {46} (\bibinfo {year} {1955})}\BibitemShut {NoStop}%
\bibitem [{\citenamefont {Goedkoop}\ and\ \citenamefont
  {Riste}(1960)}]{GoedkoopRiste_Nature1960}%
  \BibitemOpen
  \bibfield  {author} {\bibinfo {author} {\bibfnamefont {J.~A.}\ \bibnamefont
  {Goedkoop}}\ and\ \bibinfo {author} {\bibfnamefont {T.}~\bibnamefont
  {Riste}},\ }\bibfield  {title} {\bibinfo {title} {Neutron diffraction study
  of antiferromagnetic spin waves in $\alpha$-ferric oxide},\ }\href
  {https://doi.org/10.1038/185450a0} {\bibfield  {journal} {\bibinfo  {journal}
  {Nature}\ }\textbf {\bibinfo {volume} {185}},\ \bibinfo {pages} {450}
  (\bibinfo {year} {1960})}\BibitemShut {NoStop}%
\bibitem [{\citenamefont {Riste}\ and\ \citenamefont
  {Wanic}(1961)}]{Riste_1961}%
  \BibitemOpen
  \bibfield  {author} {\bibinfo {author} {\bibfnamefont {T.}~\bibnamefont
  {Riste}}\ and\ \bibinfo {author} {\bibfnamefont {A.}~\bibnamefont {Wanic}},\
  }\bibfield  {title} {\bibinfo {title} {A neutron diffraction study of spin
  fluctuations in $\alpha$-Fe$_{2}$O$_{3}$},\ }\href
  {https://doi.org/10.1016/0022-3697(61)90198-6} {\bibfield  {journal}
  {\bibinfo  {journal} {Journal of Physics and Chemistry of Solids}\ }\textbf
  {\bibinfo {volume} {17}},\ \bibinfo {pages} {318} (\bibinfo {year}
  {1961})}\BibitemShut {NoStop}%
\bibitem [{\citenamefont {Frikkee}(1966)}]{Frikkee_PhysicaB1966}%
  \BibitemOpen
  \bibfield  {author} {\bibinfo {author} {\bibfnamefont {E.}~\bibnamefont
  {Frikkee}},\ }\bibfield  {title} {\bibinfo {title} {Inelastic scattering of
  neutrons by spin waves in f.c.c. cobalt},\ }\href
  {https://doi.org/10.1016/0031-8914(66)90175-3} {\bibfield  {journal}
  {\bibinfo  {journal} {Physica}\ }\textbf {\bibinfo {volume} {32}},\ \bibinfo
  {pages} {2149} (\bibinfo {year} {1966})}\BibitemShut {NoStop}%
\bibitem [{\citenamefont {Ferguson}\ and\ \citenamefont
  {S\'{a}enz}(1967)}]{Ferguson_PhysRev1967}%
  \BibitemOpen
  \bibfield  {author} {\bibinfo {author} {\bibfnamefont {G.~A.}\ \bibnamefont
  {Ferguson}}\ and\ \bibinfo {author} {\bibfnamefont {A.~W.}\ \bibnamefont
  {S\'{a}enz}},\ }\bibfield  {title} {\bibinfo {title} {Scattering of polarized
  neutrons by spin waves in magnetite and yttrium iron garnet},\ }\href
  {https://doi.org/10.1103/physrev.156.632} {\bibfield  {journal} {\bibinfo
  {journal} {Physical Review}\ }\textbf {\bibinfo {volume} {156}},\ \bibinfo
  {pages} {632} (\bibinfo {year} {1967})}\BibitemShut {NoStop}%
\bibitem [{\citenamefont {Riste}\ \emph {et~al.}(1965)\citenamefont {Riste},
  \citenamefont {Shirane}, \citenamefont {Alperin},\ and\ \citenamefont
  {Pickart}}]{Riste_1965}%
  \BibitemOpen
  \bibfield  {author} {\bibinfo {author} {\bibfnamefont {T.}~\bibnamefont
  {Riste}}, \bibinfo {author} {\bibfnamefont {G.}~\bibnamefont {Shirane}},
  \bibinfo {author} {\bibfnamefont {H.~A.}\ \bibnamefont {Alperin}},\ and\
  \bibinfo {author} {\bibfnamefont {S.~J.}\ \bibnamefont {Pickart}},\
  }\bibfield  {title} {\bibinfo {title} {Spin-wave scattering of polarized
  neutrons from nickel and cobalt},\ }\href {https://doi.org/10.1063/1.1714107}
  {\bibfield  {journal} {\bibinfo  {journal} {Journal of Applied Physics}\
  }\textbf {\bibinfo {volume} {36}},\ \bibinfo {pages} {1076} (\bibinfo {year}
  {1965})}\BibitemShut {NoStop}%
\bibitem [{\citenamefont {Shirane}\ \emph {et~al.}(1965)\citenamefont
  {Shirane}, \citenamefont {Nathans}, \citenamefont {Steinsvoll}, \citenamefont
  {Alperin},\ and\ \citenamefont {Pickart}}]{Shirane_PRL1965}%
  \BibitemOpen
  \bibfield  {author} {\bibinfo {author} {\bibfnamefont {G.}~\bibnamefont
  {Shirane}}, \bibinfo {author} {\bibfnamefont {R.}~\bibnamefont {Nathans}},
  \bibinfo {author} {\bibfnamefont {O.}~\bibnamefont {Steinsvoll}}, \bibinfo
  {author} {\bibfnamefont {H.~A.}\ \bibnamefont {Alperin}},\ and\ \bibinfo
  {author} {\bibfnamefont {S.~J.}\ \bibnamefont {Pickart}},\ }\bibfield
  {title} {\bibinfo {title} {Measurement of the magnon dispersion relation of
  iron},\ }\href {https://doi.org/10.1103/physrevlett.15.146} {\bibfield
  {journal} {\bibinfo  {journal} {Physical Review Letters}\ }\textbf {\bibinfo
  {volume} {15}},\ \bibinfo {pages} {146} (\bibinfo {year} {1965})}\BibitemShut
  {NoStop}%
\bibitem [{\citenamefont {Alperin}\ \emph {et~al.}(1967)\citenamefont
  {Alperin}, \citenamefont {Steinsvoll}, \citenamefont {Nathans},\ and\
  \citenamefont {Shirane}}]{Alperin_PhysRev1967}%
  \BibitemOpen
  \bibfield  {author} {\bibinfo {author} {\bibfnamefont {H.~A.}\ \bibnamefont
  {Alperin}}, \bibinfo {author} {\bibfnamefont {O.}~\bibnamefont {Steinsvoll}},
  \bibinfo {author} {\bibfnamefont {R.}~\bibnamefont {Nathans}},\ and\ \bibinfo
  {author} {\bibfnamefont {G.}~\bibnamefont {Shirane}},\ }\bibfield  {title}
  {\bibinfo {title} {Magnon scattering of polarized neutrons by the diffraction
  method: Measurements on magnetite},\ }\href
  {https://doi.org/10.1103/physrev.154.508} {\bibfield  {journal} {\bibinfo
  {journal} {Physical Review}\ }\textbf {\bibinfo {volume} {154}},\ \bibinfo
  {pages} {508} (\bibinfo {year} {1967})}\BibitemShut {NoStop}%
\bibitem [{Sup()}]{Supplementary}%
  \BibitemOpen
  \href@noop {} {\bibinfo {title} {See supplementary information for details of
  analysis, fitting, and additional data presentation.}}\BibitemShut {Stop}%
\bibitem [{\citenamefont {Wang}\ \emph {et~al.}(2018)\citenamefont {Wang},
  \citenamefont {Xu}, \citenamefont {Sun},\ and\ \citenamefont
  {Xia}}]{Wang_PRM2018}%
  \BibitemOpen
  \bibfield  {author} {\bibinfo {author} {\bibfnamefont {Y.-Y.}\ \bibnamefont
  {Wang}}, \bibinfo {author} {\bibfnamefont {S.}~\bibnamefont {Xu}}, \bibinfo
  {author} {\bibfnamefont {L.-L.}\ \bibnamefont {Sun}},\ and\ \bibinfo {author}
  {\bibfnamefont {T.-L.}\ \bibnamefont {Xia}},\ }\bibfield  {title} {\bibinfo
  {title} {Quantum oscillations and coherent interlayer transport in a new
  topological Dirac semimetal candidate YbMnSb$_2$},\ }\href
  {https://doi.org/10.1103/PhysRevMaterials.2.021201} {\bibfield  {journal}
  {\bibinfo  {journal} {Phys. Rev. Mater.}\ }\textbf {\bibinfo {volume} {2}},\
  \bibinfo {pages} {021201} (\bibinfo {year} {2018})}\BibitemShut {NoStop}%
\bibitem [{\citenamefont {Arnold}\ \emph {et~al.}(2014)\citenamefont {Arnold},
  \citenamefont {Bilheux}, \citenamefont {Borreguero}, \citenamefont {Buts},
  \citenamefont {Campbell}, \citenamefont {Chapon}, \citenamefont {Doucet},
  \citenamefont {Draper}, \citenamefont {Leal}, \citenamefont {Gigg} \emph
  {et~al.}}]{arnold_mantid2014}%
  \BibitemOpen
  \bibfield  {author} {\bibinfo {author} {\bibfnamefont {O.}~\bibnamefont
  {Arnold}}, \bibinfo {author} {\bibfnamefont {J.-C.}\ \bibnamefont {Bilheux}},
  \bibinfo {author} {\bibfnamefont {J.}~\bibnamefont {Borreguero}}, \bibinfo
  {author} {\bibfnamefont {A.}~\bibnamefont {Buts}}, \bibinfo {author}
  {\bibfnamefont {S.~I.}\ \bibnamefont {Campbell}}, \bibinfo {author}
  {\bibfnamefont {L.}~\bibnamefont {Chapon}}, \bibinfo {author} {\bibfnamefont
  {M.}~\bibnamefont {Doucet}}, \bibinfo {author} {\bibfnamefont
  {N.}~\bibnamefont {Draper}}, \bibinfo {author} {\bibfnamefont {R.~F.}\
  \bibnamefont {Leal}}, \bibinfo {author} {\bibfnamefont {M.}~\bibnamefont
  {Gigg}}, \emph {et~al.},\ }\bibfield  {title} {\bibinfo {title}
  {Mantid — data analysis and visualization package for neutron scattering and
  $\mu$-SR experiments},\ }\href {https://doi.org/10.1016/j.nima.2014.07.029}
  {\bibfield  {journal} {\bibinfo  {journal} {Nuclear instruments and methods
  in physics research section A: accelerators, spectrometers, detectors and
  associated equipment}\ }\textbf {\bibinfo {volume} {764}},\ \bibinfo {pages}
  {156} (\bibinfo {year} {2014})}\BibitemShut {NoStop}%
\bibitem [{\citenamefont {Savici}\ \emph {et~al.}(2022)\citenamefont {Savici},
  \citenamefont {Gigg}, \citenamefont {Arnold}, \citenamefont {Tolchenov},
  \citenamefont {Whitfield}, \citenamefont {Hahn}, \citenamefont {Zhou},\ and\
  \citenamefont {Zaliznyak}}]{savici_MDnorm2022}%
  \BibitemOpen
  \bibfield  {author} {\bibinfo {author} {\bibfnamefont {A.~T.}\ \bibnamefont
  {Savici}}, \bibinfo {author} {\bibfnamefont {M.~A.}\ \bibnamefont {Gigg}},
  \bibinfo {author} {\bibfnamefont {O.}~\bibnamefont {Arnold}}, \bibinfo
  {author} {\bibfnamefont {R.}~\bibnamefont {Tolchenov}}, \bibinfo {author}
  {\bibfnamefont {R.~E.}\ \bibnamefont {Whitfield}}, \bibinfo {author}
  {\bibfnamefont {S.~E.}\ \bibnamefont {Hahn}}, \bibinfo {author}
  {\bibfnamefont {W.}~\bibnamefont {Zhou}},\ and\ \bibinfo {author}
  {\bibfnamefont {I.~A.}\ \bibnamefont {Zaliznyak}},\ }\bibfield  {title}
  {\bibinfo {title} {Efficient data reduction for time-of-flight neutron
  scattering experiments on single crystals},\ }\href
  {https://doi.org/10.1107/S1600576722009645} {\bibfield  {journal} {\bibinfo
  {journal} {Journal of Applied Crystallography}\ }\textbf {\bibinfo {volume}
  {55}} (\bibinfo {year} {2022})}\BibitemShut {NoStop}%
\end{thebibliography}

%


\begin{widetext}
\pagebreak
\end{widetext}
\hypersetup{pageanchor=false}
\renewcommand{\thepage}{S\arabic{page}}
\setcounter{page}{1}
\renewcommand{\theequation}{S\arabic{equation}}
\setcounter{equation}{0}
\renewcommand{\thefigure}{S\arabic{figure}}
\setcounter{figure}{0}
\setcounter{table}{0}

\begin{widetext}

\section*{Supplementary Information}

\begin{center}
{\bf Spin waves in Dirac semimetal \cms\ investigated with neutrons by the diffraction method} \\
Xiao Hu, Yan Wu, Matthias D. Frontzek, Zhixiang Hu, Cedomir Petrovic, John M. Tranquada, and Igor A. Zaliznyak
correspondence to: zaliznyak@bnl.gov
\end{center}
\bigskip
\noindent{\bf This PDF file includes:}\\
Supplementary Text\\
Supplementary Tables S1, S2\\
Supplementary Figures S1-S11\\

\subsection{Projected energy-integrated intensity in neutron diffraction measurement}
\label{Model}

\begin{figure}[b]
\includegraphics[width=0.67\textwidth]{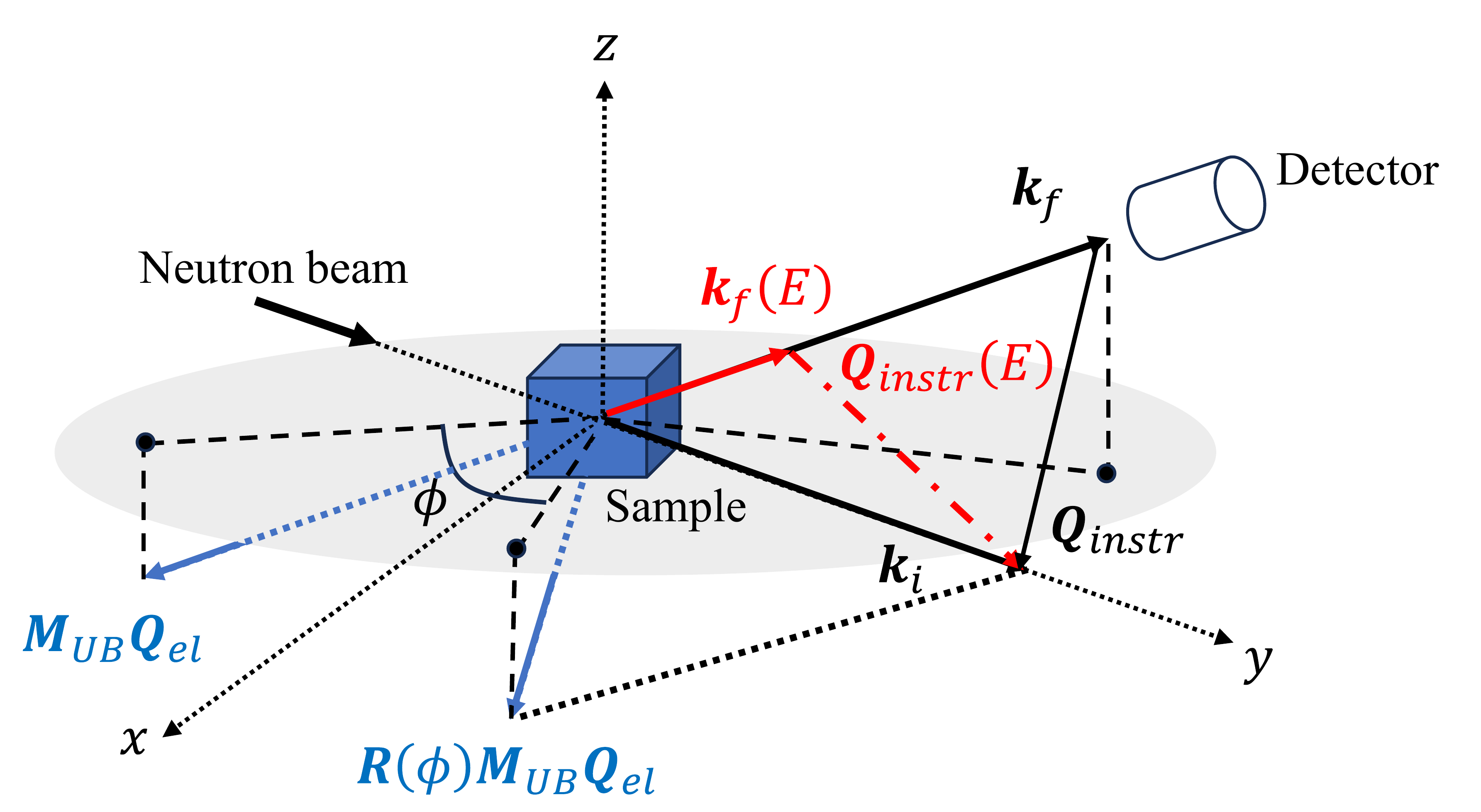}
\caption{Schematics of neutron diffraction measurement. Light grey ellipse represents the horizontal scattering plane, $(x, y, z)$ axes represent the instrument coordinates. See text for other notations.}
\label{supple1_diffraction}
\end{figure}

The schematics of diffraction measurement without scattered neutron energy analysis which illustrates our projected (energy-integrated) model intensity calculation is shown in Figure~\ref{supple1_diffraction}. $(x, y, z)$ axes represent instrument coordinate system with $y$-axis directed along the incident beam (incident neutron wave vector, $\bk_{i}$) and the vertical $z$-axis directed upwards. For a given detector, its angular position in instrument coordinates is specified by $\bk_{f} \equiv \bk_{f} (0)$, the scattered neutron wave vector for elastic scattering with the given fixed $\bk_{i}$ used for the measurement, $k_{f} = k_{i}$. The corresponding wave vector transfer, $\bQ_{instr} \equiv \bQ_{instr}(0) = \bk_{i} - \bk_{f}$, satisfies energy-momentum conservation for elastic scattering (Bragg's law; the subscript ${instr}$ indicates that sample wave vector transfer here is referred to the instrument coordinate system, along with $\bk_{i}$ and $\bk_{f}(E)$). For each sample orientation, the detector collects all neutrons scattered along the $\bk_{f}$ direction, with all different $\bk_{f}(E) \parallel \bk_{f}$ corresponding to different energy and momentum transfers, $E$ and $\bQ_{instr} (E)$, satisfying energy-momentum conservation,
\begin{equation}
\label{Eq_E_conserv}
E = E_i - E_f = \frac{\hbar^2}{2m_n}(\bk_{i}^2 - \bk_{f}^2(E)) \, ,
\end{equation}
\begin{equation}
\label{Eq_Q_conserv}
\bQ_{instr}(E) = \bk_{i} - \bk_{f}(E) ,\; \bk_{f}(E) \parallel \bk_{f} \, ,
\end{equation}
where $-\infty < E < E_i$. It follows from Eqs.~\eqref{Eq_E_conserv}, \eqref{Eq_Q_conserv}, that,
\begin{equation}
\label{Eq_Q_E}
\bQ_{instr}(E) = \bk_{i} - \frac{\bk_{f}}{k_{i}}k_{f}(E) = \bk_{i} - \left( \bk_{i} - \bQ_{instr}(0) \right) \sqrt{1 - E/E_{i}} ,
\end{equation}
which is equivalent to Eq.~(2) of the main text written in the instrument coordinate system, Fig.~\ref{supple1_diffraction}.

%
In a diffraction experiment, all neutrons measured by a given detector element positioned at some $\bk_f$ are assigned to elastic scattering with the wave vector transfer $\bQ_{instr} = \bk_i - \bk_f$ in the instrument coordinate frame. In the sample reciprocal lattice coordinates, this $\bQ_{instr}$ corresponds to wave vector transfer, $(H,K,L) = \bQ(0) \equiv \bQ_{el}$, which depends on sample orientation specified by the rotation angle, $\phi$, around the vertical axis, $z$ (Fig.~\ref{supple1_diffraction}). Therefore, in order to calculate the projected scattering intensity at an arbitrary reciprocal space point, $(H,K,L)$, we need to determine the sample rotation angle, $\phi$, at which $\bQ_{el}$ transferred to the instrument coordinates coincides with $\bQ_{instr}$ (for the given $\bk_i$ used in the neasurement). That is, we need to find $\phi$ at which this $(H,K,L)$ point was measured. Note, this requires that scattering triangle can be closed for some $\bk_f = \bk_i - \bQ_{instr}$ (there might or might not be a detector element at this $\bk_f$).

The transformation of wave vectors from the sample reciprocal space to the instrument coordinates (and to the absolute units of \AA$^{-1}$) is encoded in the UB matrix, $\bM_{UB}$, obtained from sample alignment and defined relative to sample rotation $\phi  = 0$. Additional sample rotation by an angle $\phi$ required to satisfy elastic scattering condition is encoded in the rotation matrix, $\bR (\phi)$. Then,
\begin{equation}
\label{Eq_UB_R_Qel}
\bQ_{instr} = \bR(\phi) \bM_{UB} \bQ_{el} ,
\end{equation}
determines the sample rotation angle, $\phi$, for a given $\bQ_{el} = (H, K, L)$ and $\bk_i$. With this $\phi$, the corresponding transformation from sample to instrument coordinates for all $\bQ_{}(E)$ contributing to the measured intensity at $(H, K, L)$ is,
\begin{equation}
\label{Eq_UB_R_Q_E}
\bQ_{instr}(E) = \bR(\phi) \bM_{UB} \bQ_{}(E) ,\; \bk_{f}(E) = \bk_{i} - \bR(\phi) \bM_{UB} \bQ_{} (E) .
\end{equation}%
Transforming back to sample reciprocal lattice coordinates and using Eq.~\eqref{Eq_Q_E} we obtain,
\begin{equation}
\label{Eq_Q_E_samp}
\bQ_{} (E) =
\bM^{-1}_{UB} \bR^{-1}(\phi) \left( \bk_{i} - \bk_{f}(E) \right) =
\bM^{-1}_{UB} \bR^{-1}(\phi) \bk_{i} \left(1-\sqrt{1-E/E_i} \right) + \bQ_{el} \sqrt{1 - E/E_{i}} .
\end{equation}
This is Eq.~(2) of the main text written in the sample reciprocal space coordinates (with $\bk_i$ moved to sample reciprocal space).

In order to obtain the total projected scattering intensity from processes with all possible energy transfers, $E$, for a given $\bQ_{el} = (H, K, L)$, we substitute the corresponding $\bQ(E)$ from Eq.~\eqref{Eq_Q_E_samp} into a model cross-section, $\frac{d^2 \sigma(\bQ, E)}{dE d\Omega}$, and integrate in $dE$,
\begin{equation}
\label{Eq_I_Qel}
I({\bQ_{el}})
= \tilde{A} \int_{-\infty}^{E_{i}} \frac{d^2 \sigma(\bQ(E), E)}{dE d\Omega} \,dE \
= A \int_{-\infty}^{E_{i}} \dfrac{k_{f}(E)}{k_{i}} S(\bQ(E)\emph{}, E) \,dE ,
\end{equation}
where $S(\bq, E)$ is the dynamical spin correlation function (in the sample reciprocal space coordinates) and $A$ ($\tilde{A}$) is the normalization factor. In our model, we used spin-wave cross-section of $J_{1} - J_{2} - J_{c}$ Heisenberg model described in detail in Refs.~ \onlinecite{Sapkota_2020,Hu_2023}, which was found to describe well similar quasi-2D Dirac semimetal systems, YbMnBi$_2$ and YbMnSb$_2$.

\subsection{Details of the fitting procedure and results}
\label{Fitting}

\begin{table}[b!]
\renewcommand{\thetable}{S\arabic{table}}
\caption{\label{Tabparameter} Spin wave parameters obtained by fitting \cms\ diffraction data at different temperatures. Values in parentheses represent uncertainties with $95\%$ fitting confidence or indicate the parameter being fixed.}
\begin{ruledtabular}
\begin{tabular}{cccccccc}
&\mbox{10 K}&\mbox{50 K}&\mbox{100 K}&\mbox{150 K}&\mbox{200 K}&\mbox{250 K}&\mbox{300 K}\\
\hline
$SJ_1$ (meV) & $25.2$ (fixed) & $23.1(22)$  & $24.4(18)$   & $24.6(22)$   & $26.2(34)$    & $25.7(76)$    & $25.2$ (fixed)\\
$SJ_2$ (meV) & $8.6$ (fixed) & $8.0(14)$   & $8.2(12)$    & $8.3(14)$    & $9.2(22)$     & $8.6(50)$     & $8.6$  (fixed)\\
$SJ_c$ (meV) & $-0.190(8)$ & $-0.203(20)$   & $-0.166(14)$  & $-0.144(14)$  & $-0.112(15)$   & $-0.091(28)$   & $-0.059(6)$\\
$SD$ (meV)   & $-0.086(5)$ & $-0.084(9)$    & $-0.069(7)$ & $-0.051(5)$  & $-0.036(6)$  & $-0.034(9)$  & $0$ (fixed)\\
$\Delta$ (meV)   & $5.9(2)$    & $5.6(4)$    & $5.2(3)$    & $4.5(3)$    & $3.9(4)$     & $3.7(8)$    & $0$\\
$\gamma$ (meV) & $0.9$ (fixed)   & $0.9$ (fixed)   & $0.9(1)$    & $1.0(2)$    & $2.0(2)$     & $6.3(3)$   & $11.8(3)$\\
\end{tabular}
\end{ruledtabular}
\end{table}


In order to account for the effects of the instrumental wave vector resolution and the finite bin size to which the data is re-binned, the calculated scattering cross-section [dynamical spin correlation function, $S(\bq,E)$] was convoluted with the appropriate experimental resolution function as described in the supplementary information of Ref.~{\onlinecite{Hu_2023}}. For fitting, we used three different $L$-slices, averaged for $L \in [-0.6,-0.4], [-0.35,-0.15]$, and $[-0.1,0.1]$, and with the binning size of $(\pm0.01, \pm0.01)$ in $(H,K)$, which provide sufficient intensity for fitting while minimizing the resolution effects. The instrumental \bQ-resolution was evaluated by fitting $(0,-1,0)$ Bragg peak at 10 K binned on a much finer grid, $(\Delta H, \Delta K, \Delta L) = (0.0025, 0.005, 0.06)$, to Gaussian function.

Figures~\ref{supp2_10K} through \ref{supp8_300K} present the three constant-$L$ slices of the data for $L \in [-0.6,-0.4], [-0.35,-0.15]$, and $[-0.1,0.1]$ [panels (a--c)], which were fitted simultaneously to our model described above in order to refine the spin wave parameters for different temperatures we measured, which are listed in Table~\ref{Tabparameter}. Panels (d--f) in the figures show the corresponding calculated model intensity, while the line cuts of the data are compared with the model intensity in panels (g--l). A weak band of intensity, which is visible at smaller wave vectors, around $(H,K)\sim(-0.3,-0.3)$, in the simulated intensity [panels (d--f)] but also in the data at higher temperatures, 200~K and above, arises from the spin-wave spectra on the neutron energy-gain side (negative energy transfers, $E<0$). At low $T$ this intensity is greatly suppressed due to the condition of detailed balance; it becomes more visible when the temperature increases.

For 100 K (Fig.~\ref{supp4_100K}), 150 K (Fig.~\ref{supp5_150K}), 200 K (Fig.~\ref{supp6_200K}), 250 K (Fig.~\ref{supp7_250K}), the data sets corresponding to $L \in [-0.6,-0.4], [-0.35,-0.15]$, and $[-0.1,0.1]$ were fitted simultaneously varying all six parameters listed in Table~\ref{Tabparameter}. For 300 K (Fig.~\ref{supp8_300K}), the projected spin-wave spectra do not allow to confidently determine $SJ_1$ and $SJ_2$ and the spin gap (Q-gap), which is closed. Hence, we fixed $SJ_1$ and $SJ_2$ to the average values obtained from 100~K to 250~K fitting and fixed the gap to zero, setting $SD = 0$. The inter-layer exchange parameter $SJ_c$ was obtained by fitting the three $L$ slices simultaneously.

For 10~K (Fig.~\ref{supp2_10K}), the scattering intensity is rather weak and statistics is insufficient to obtain a reliable fit varying all parameters. Firstly, the spin-wave spectral damping $\gamma$ cannot be confidently resolved (unlike in 100 K and 150 K data) and was fixed to the 100~K value. Secondly, varying all three exchange couplings yields $SJ_1$ and $SJ_2$ within an $\sim 10\%$ error bar from those reported in Table~\ref{Tabparameter}, but an unreasonably, 1.5 to 2 times larger $SJ_c$. From all the different fit procedures we examined for 10~K data, the best fit is obtained when $SJ_1$ and $SJ_2$ are fixed to the average values of 100~K to 250~K fits and $SJ_c$ and $SD$ are fitted. These results are reported in Table~\ref{Tabparameter} as the best fit parameters. The fitted parameters for all temperatures are summarized in Table.~\ref{Tabparameter}.

%
\begin{figure}[h]
\includegraphics[width=0.75\textwidth]{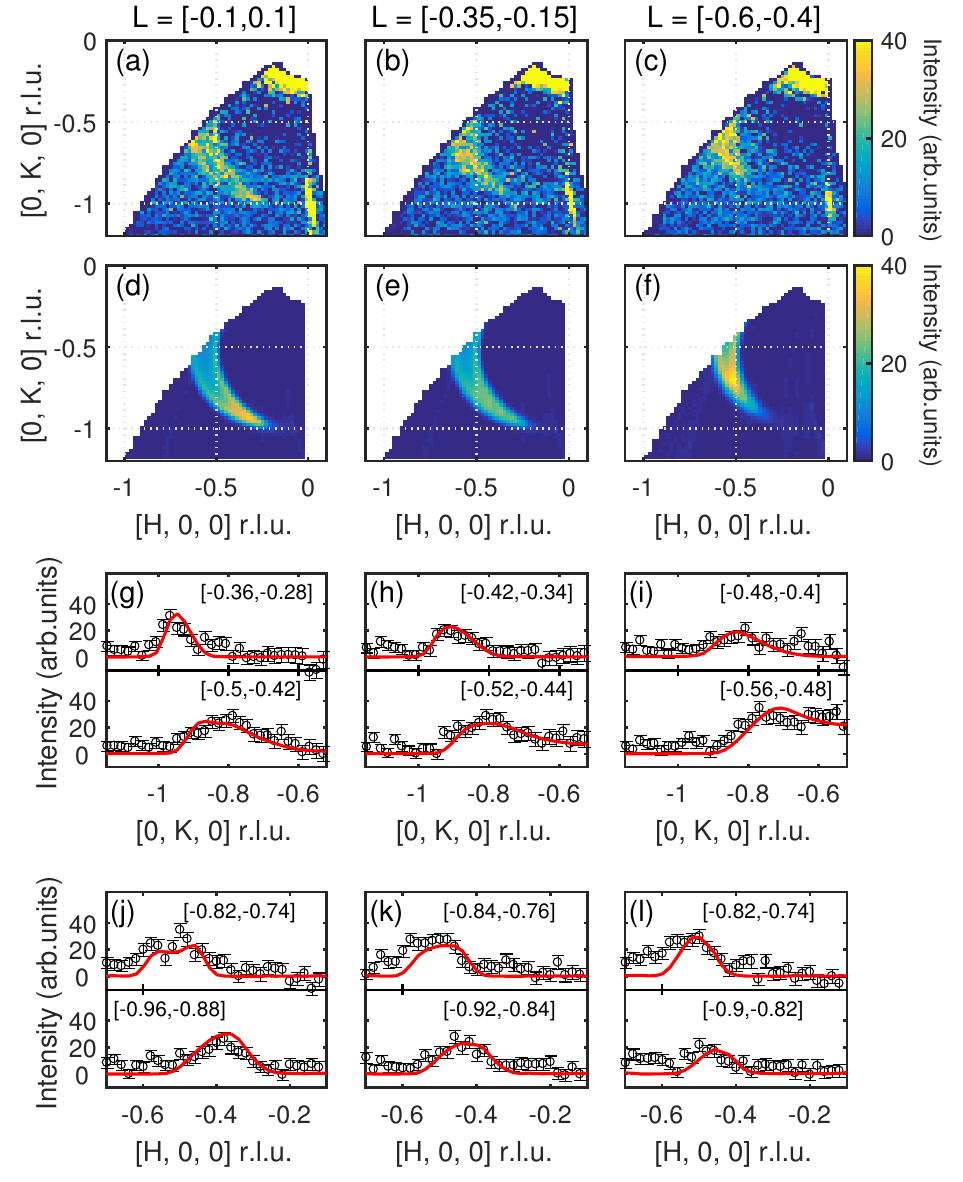}
\caption{Projected spin wave spectra of \cms\ measured by diffraction at $T = 10$~K. The data was averaged in the range $L = [-0.1,0.1]$ (a), $[-0.35,-0.15]$ (b), and $[-0.6,-0.4]$ (c). The direct beam plus constant background intensity was evaluated by fitting to a 2D Gaussian profile and subtracted from the data. (d--f) are the corresponding fitted spectra using the parameters in Table~\ref{Tabparameter}. The constant-$H$ cuts (g--i) and constant-$K$ cuts (j--l) show direct comparison of fits with the data. The filled circles with error bars representing one standard deviation are the measured intensity and the red solid lines are the fit. }
\label{supp2_10K}
\end{figure}

\begin{figure}[h]
\includegraphics[width=0.75\textwidth]{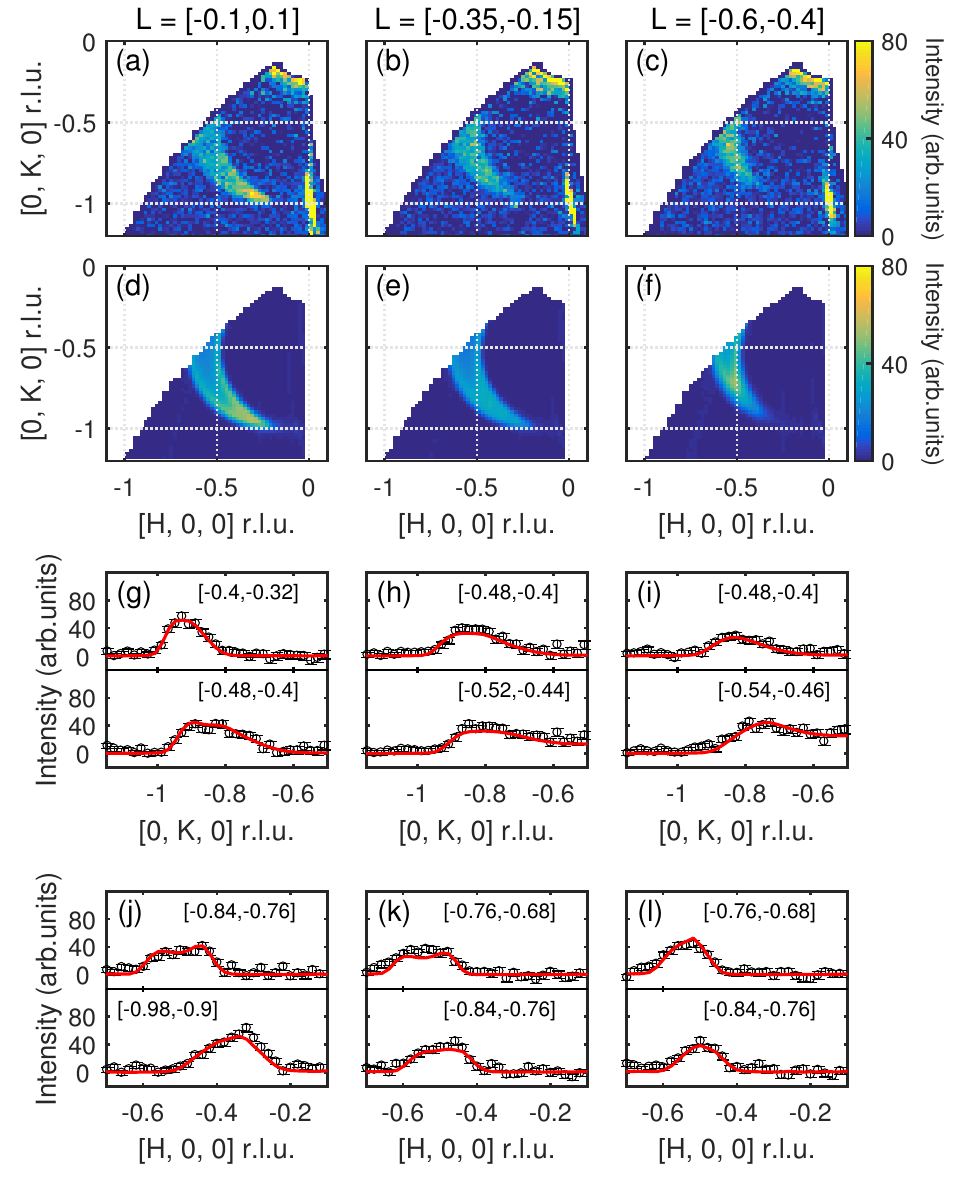}
\caption{Projected spin wave spectra of \cms\ measured by diffraction at $T = 50$~K. The data was averaged in the range $L = [-0.1,0.1]$ (a), $[-0.35,-0.15]$ (b), and $[-0.6,-0.4]$ (c). The direct beam plus constant background intensity was evaluated by fitting to a 2D Gaussian profile and subtracted from the data. (d--f) are the corresponding fitted spectra using the parameters in Table~\ref{Tabparameter}. The constant-$H$ cuts (g--i) and constant-$K$ cuts (j--l) show direct comparison of fits with the data. The filled circles with error bars representing one standard deviation are the measured intensity and the red solid lines are the fit. }
\label{supp3_50K}
\end{figure}

\begin{figure}[h]
\includegraphics[width=0.75\textwidth]{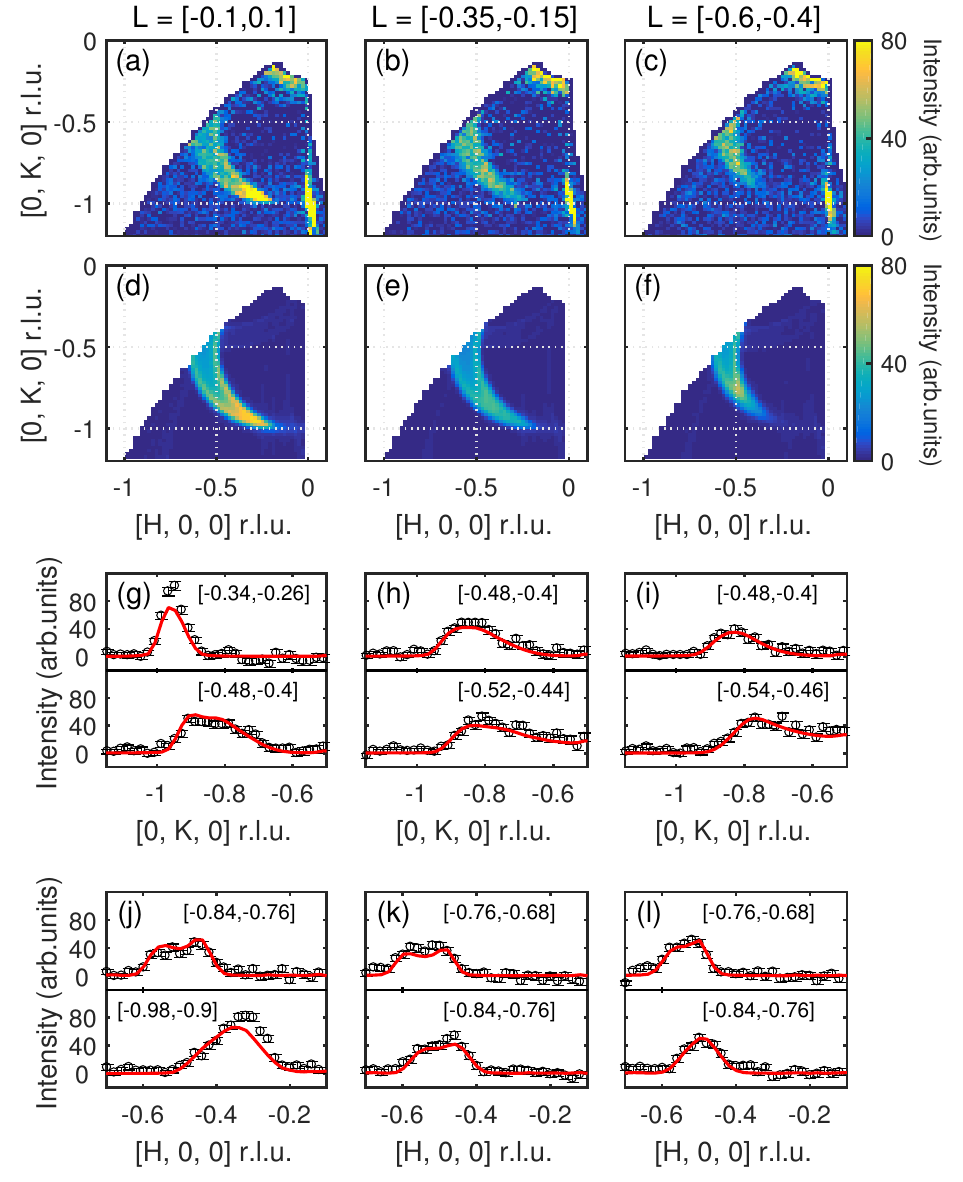}
\caption{Projected spin wave spectra of \cms\ measured by diffraction at $T = 100$~K. The data was averaged in the range $L = [-0.1,0.1]$ (a), $[-0.35,-0.15]$ (b), and $[-0.6,-0.4]$ (c). The direct beam plus constant background intensity was evaluated by fitting to a 2D Gaussian profile and subtracted from the data. (d--f) are the corresponding fitted spectra using the parameters in Table~\ref{Tabparameter}. The constant-$H$ cuts (g--i) and constant-$K$ cuts (j--l) show direct comparison of fits with the data. The filled circles with error bars representing one standard deviation are the measured intensity and the red solid lines are the fit. }
\label{supp4_100K}
\end{figure}

\begin{figure}[h]
\includegraphics[width=0.75\textwidth]{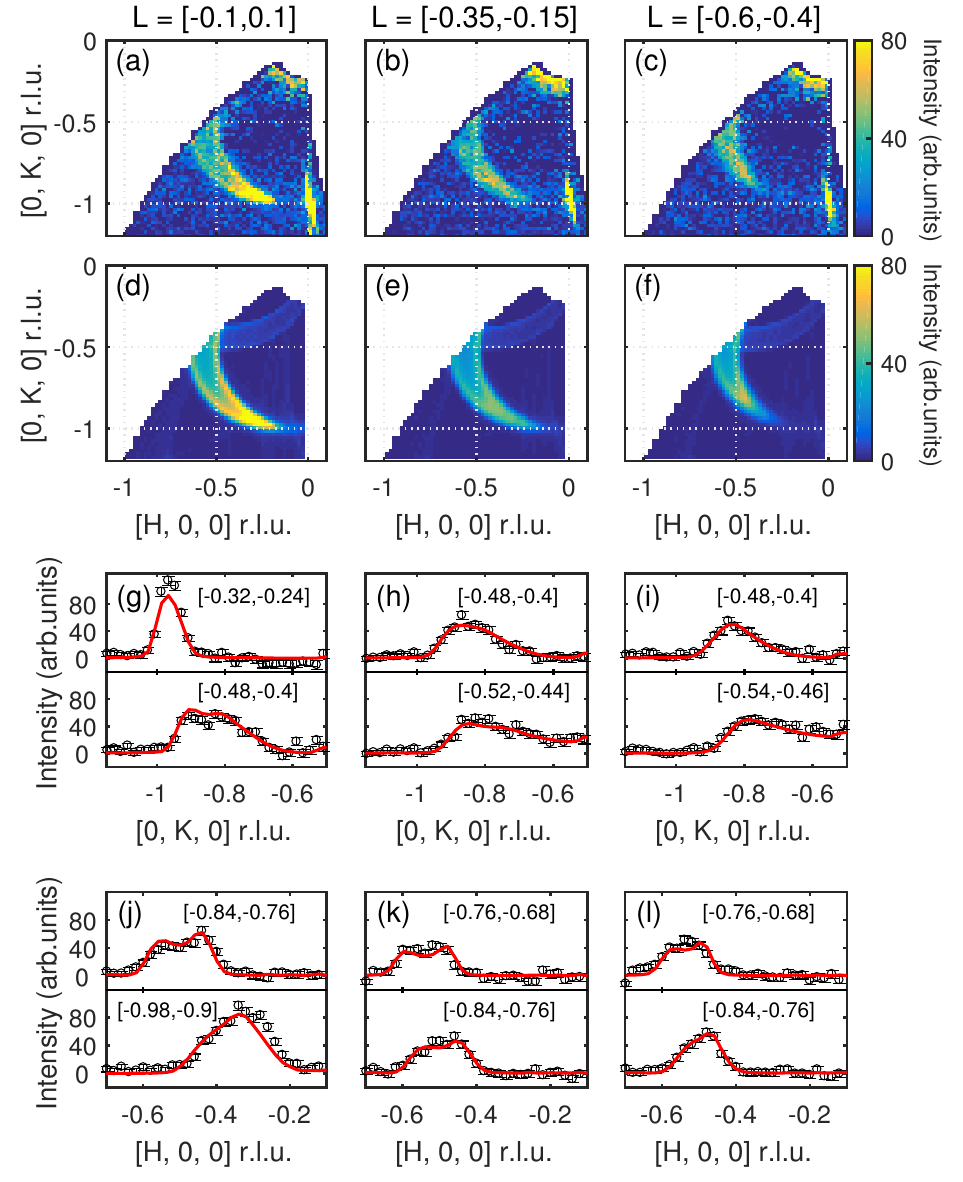}
\caption{Projected spin wave spectra of \cms\ measured by diffraction at $T = 150$~K. The data was averaged in the range $L = [-0.1,0.1]$ (a), $[-0.35,-0.15]$ (b), and $[-0.6,-0.4]$ (c). The direct beam plus constant background intensity was evaluated by fitting to a 2D Gaussian profile and subtracted from the data. (d--f) are the corresponding fitted spectra using the parameters in Table~\ref{Tabparameter}. The constant-$H$ cuts (g--i) and constant-$K$ cuts (j--l) show direct comparison of fits with the data. The filled circles with error bars representing one standard deviation are the measured intensity and the red solid lines are the fit. }
\label{supp5_150K}
\end{figure}

\begin{figure}[h]
\includegraphics[width=0.75\textwidth]{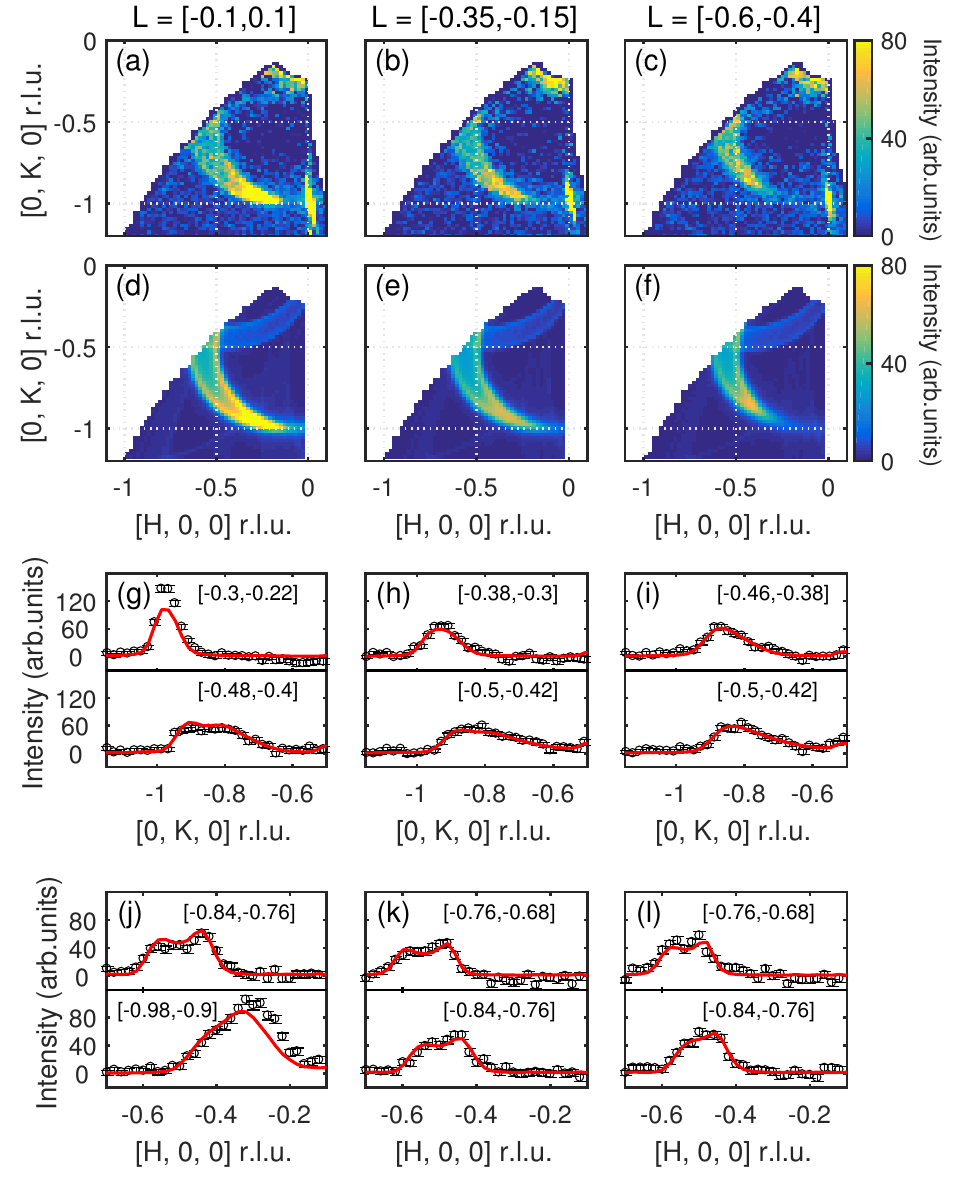}
\caption{Projected spin wave spectra of \cms\ measured by diffraction at $T = 200$~K. The data was averaged in the range $L = [-0.1,0.1]$ (a), $[-0.35,-0.15]$ (b), and $[-0.6,-0.4]$ (c). The direct beam plus constant background intensity was evaluated by fitting to a 2D Gaussian profile and subtracted from the data. (d--f) are the corresponding fitted spectra using the parameters in Table~\ref{Tabparameter}. The constant-$H$ cuts (g--i) and constant-$K$ cuts (j--l) show direct comparison of fits with the data. The filled circles with error bars representing one standard deviation are the measured intensity and the red solid lines are the fit. }
\label{supp6_200K}
\end{figure}

\begin{figure}[h]
\includegraphics[width=0.75\textwidth]{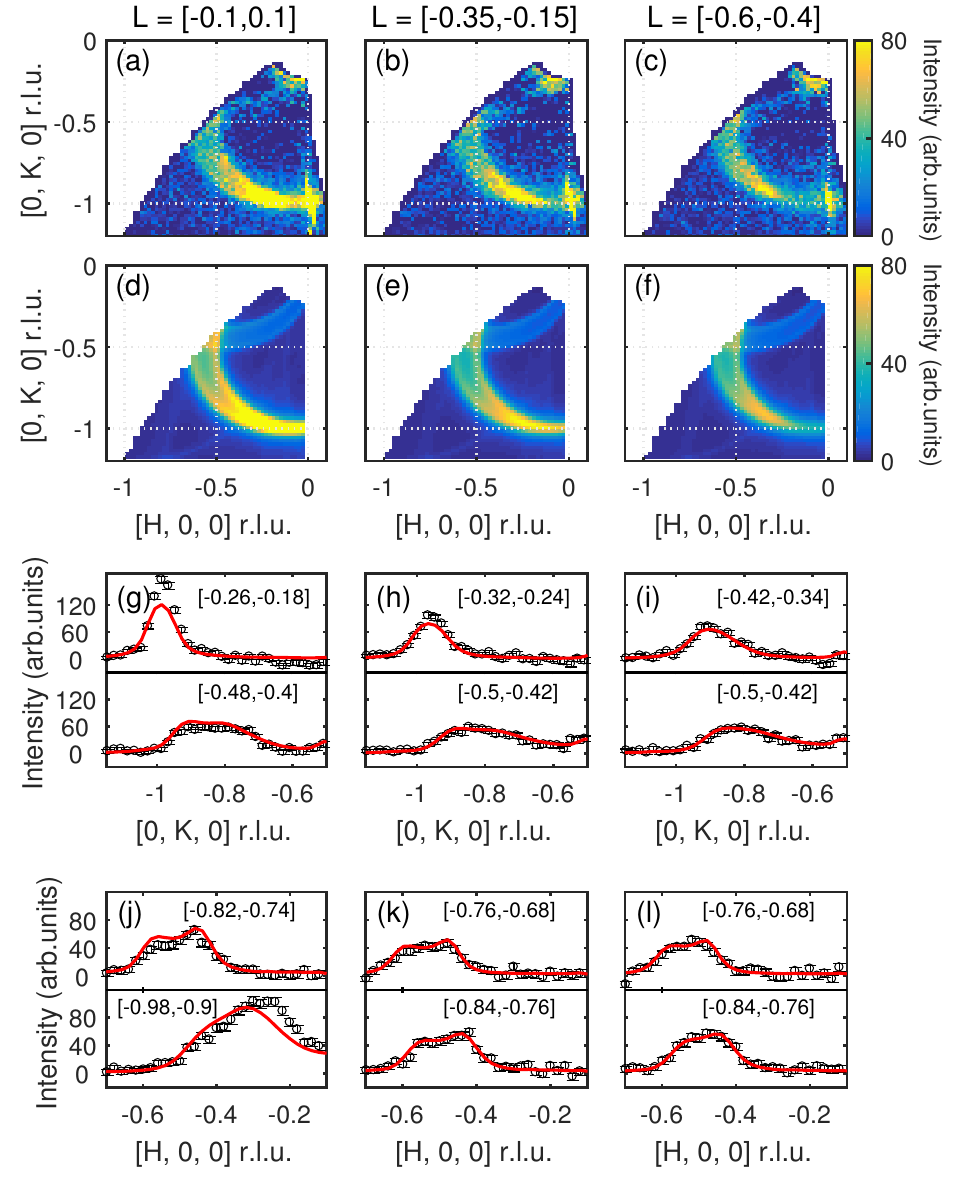}
\caption{Projected spin wave spectra of \cms\ measured by diffraction at $T = 250$~K. The data was averaged in the range $L = [-0.1,0.1]$ (a), $[-0.35,-0.15]$ (b), and $[-0.6,-0.4]$ (c). The direct beam plus constant background intensity was evaluated by fitting to a 2D Gaussian profile and subtracted from the data. (d--f) are the corresponding fitted spectra using the parameters in Table~\ref{Tabparameter}. The constant-$H$ cuts (g--i) and constant-$K$ cuts (j--l) show direct comparison of fits with the data. The filled circles with error bars representing one standard deviation are the measured intensity and the red solid lines are the fit. }
\label{supp7_250K}
\end{figure}

\begin{figure}[h]
\includegraphics[width=0.75\textwidth]{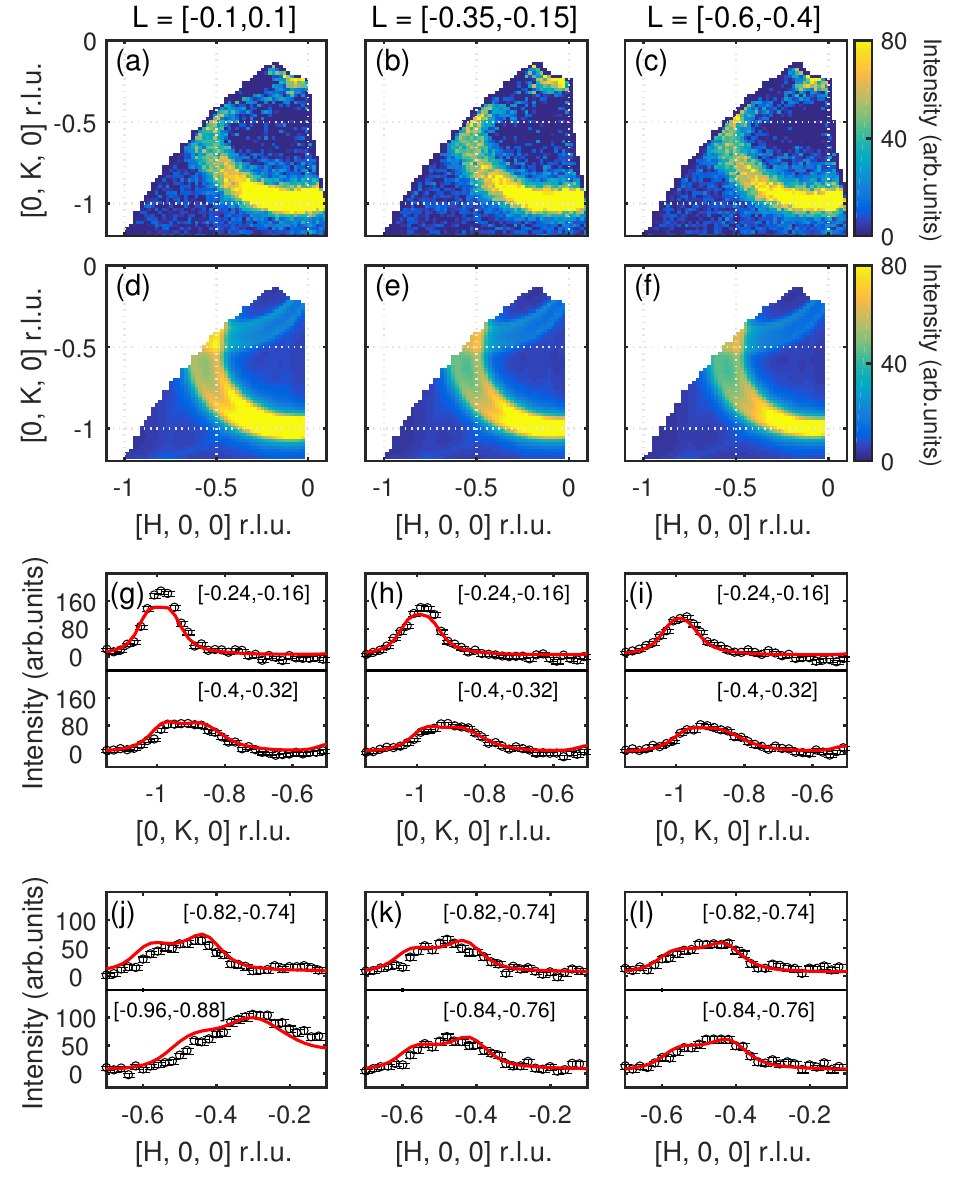}
\caption{Projected spin wave spectra of \cms\ measured by diffraction at $T = 300$~K. The data was averaged in the range $L = [-0.1,0.1]$ (a), $[-0.35,-0.15]$ (b), and $[-0.6,-0.4]$ (c). The direct beam plus constant background intensity was evaluated by fitting to a 2D Gaussian profile and subtracted from the data. (d--f) are the corresponding fitted spectra using the parameters in Table~\ref{Tabparameter}. The constant-$H$ cuts (g--i) and constant-$K$ cuts (j--l) show direct comparison of fits with the data. The filled circles with error bars representing one standard deviation are the measured intensity and the red solid lines are the fit. }
\label{supp8_300K}
\end{figure}

\subsection{Spin-wave spectral damping}
\label{damping}

\begin{table}[b!]
\renewcommand{\thetable}{S\arabic{table}}
\caption{\label{Tabpar_50K} Spin wave parameters obtained by fitting \cms\ diffraction data at 50 K to model with different parameters fixed. Values in parentheses represent uncertainties with $95\%$ fitting confidence or indicate the parameter being fixed.}
\begin{ruledtabular}
\begin{tabular}{ccccc}
            &\mbox{Model 1} &\mbox{Model 2} &\mbox{Model 3} &\mbox{Model 4} \\
\hline
$SJ_1$ (meV) & $25.1(5)$    & $23.1(22)$    & $22.3(24)$    & $25.2$ (fixed)\\
$SJ_2$ (meV) & $8.5(3)$     & $8.0(14)$     & $7.5(16)$     & $8.6$ (fixed) \\
$SJ_c$ (meV) & $-0.168(4)$  & $-0.203(20)$  & $-0.215(24)$  & $-0.179(5)$   \\
$SD$ (meV)   & $-0.072(2)$  & $-0.084(9)$   & $-0.099(12)$  & $-0.070(3)$   \\
$\Delta$ (meV) & $5.4(1)$   & $5.6(4)$      & $5.9(5)$      & $5.3(1)$      \\
$\gamma$ (meV) & $0.11$ (fixed) & $0.9$ (fixed) & $1.7(2)$  & $0.9$ (fixed) \\
$\chi^2$       & $0.81$     & $0.72$        & $0.71$        & $0.75$        \\
\end{tabular}
\end{ruledtabular}
\end{table}

\vspace{-0.5em}
A rather surprising result of this study is that our model fitting allows to resolve the spin-wave spectral damping parameter, $\gamma$, from diffraction data measured without energy analysis. It is therefore important to test the reliability of such model fitting in distinguishing the damping parameter, $\gamma$. As shown in Fig.~\ref{supp9} for the 100~K data, we tested the resolution limit of the sensitivity to damping parameter by fixing it to be $\gamma = 0.11$~meV. Compared with Fig.~\ref{supp4_100K} (optimized damping parameter $\gamma$), the $\gamma = 0.11$ meV fit in Fig.~\ref{supp9} is slightly less accurate in describing the observed spectral features of the data and, as shown in the main text, the standard deviations of the fitted interaction parameters for fixed $\gamma = 0.11$~meV are larger than those of the optimized $\gamma$ fitting. The quality of the optimized $\gamma$ fitting ($\chi^{2} = 0.98$) is also consistently better than that for fixed $\gamma = 0.11$ meV ($\chi^{2}=1.06$). Yet, the improvement is not significant, so the resolution limitation puts the damping parameter in the range $0.11 \lesssim \gamma \lesssim 0.90$~meV.

We further explored how the noise in the lower-intensity data at low temperatures impacts the $\gamma$ refinement by fitting the 50~K data with fixed $\gamma = 0.11$~meV (Fig.~\ref{supp10}) and with the $\gamma$ freely optimized (Fig.~\ref{supp11}). The results are reported in Table~\ref{Tabpar_50K}. While the $\chi^2$ values are less than 1 for all fits, there is a consistent increase when the damping is neglected (for $\gamma = 0.11$~meV), which can be taken as an indication that small damping is indeed present, similar to the 100 K data.

\begin{figure}[h]
\includegraphics[width=0.75\textwidth]{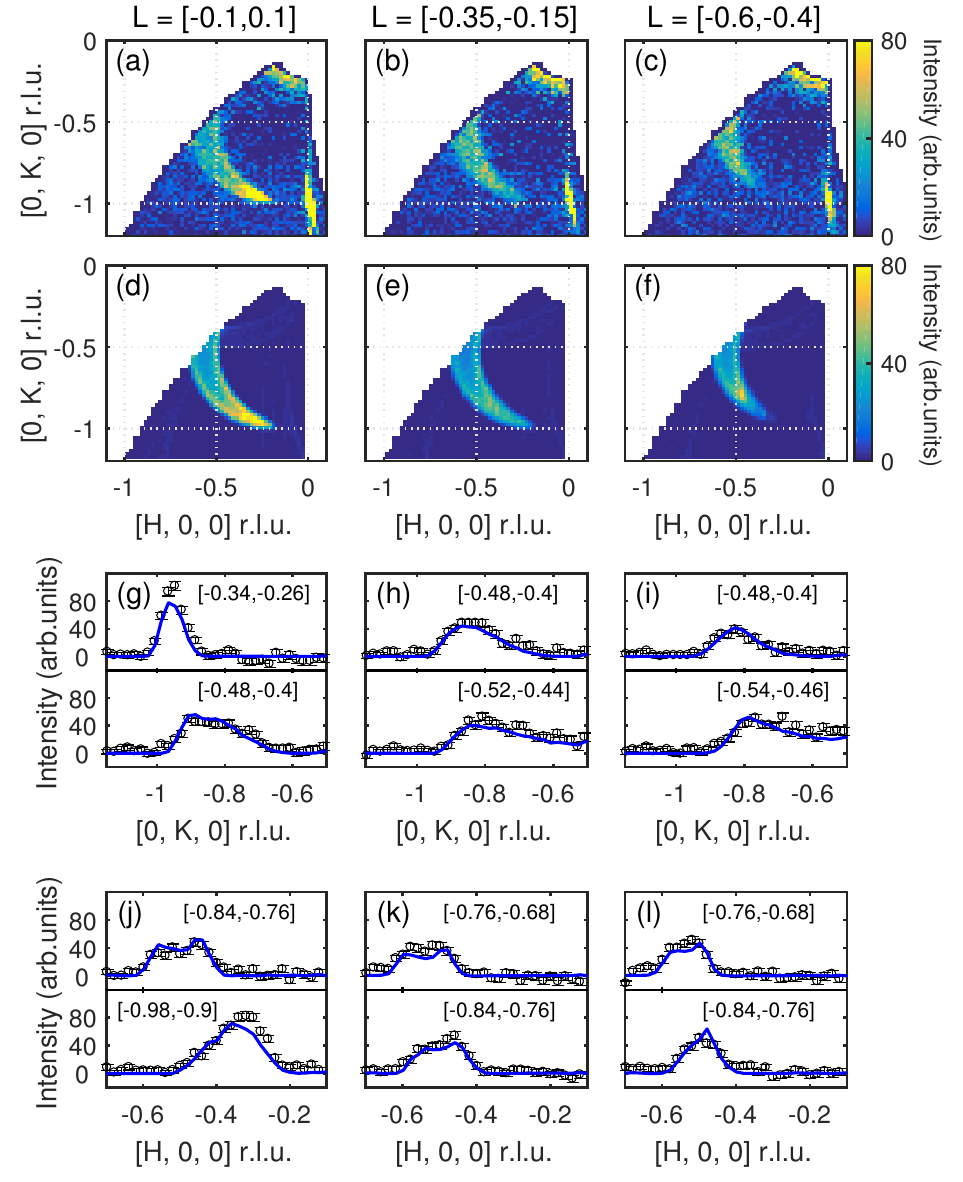}
\caption{Projected spin wave spectra of \cms\ measured by diffraction at $T = 100$~K. The data was averaged in the range $L = [-0.1,0.1]$ (a), $[-0.35,-0.15]$ (b), and $[-0.6,-0.4]$ (c). The direct beam plus constant background intensity was evaluated by fitting to a 2D Gaussian profile and subtracted from the data. (d--f) are the corresponding fitted spectra using the parameters of Model 1 ($\gamma = 0.11$) in the main text. The constant-$H$ cuts (g--i) and constant-$K$ cuts (j--l) show direct comparison of fits with the data. The filled circles with error bars representing one standard deviation are the measured intensity and the blue solid lines are the fit. }
\label{supp9}
\end{figure}

\begin{figure}[h]
\includegraphics[width=0.75\textwidth]{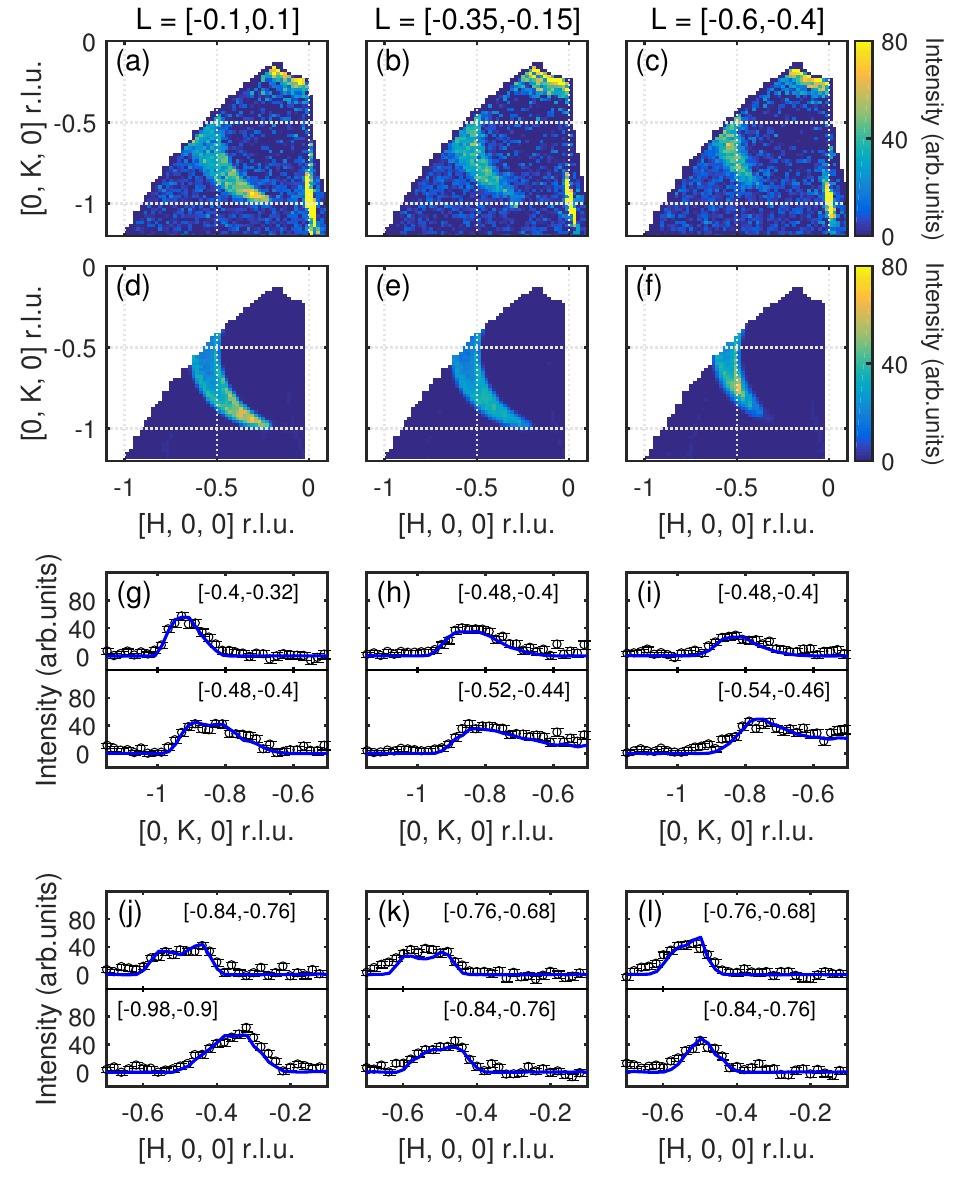}
\caption{Projected spin wave spectra of \cms\ measured by diffraction at $T = 50$~K. The data was averaged in the range $L = [-0.1,0.1]$ (a), $[-0.35,-0.15]$ (b), and $[-0.6,-0.4]$ (c). The direct beam plus constant background intensity was evaluated by fitting to a 2D Gaussian profile and subtracted from the data. (d--f) are the corresponding fitted spectra using the fixed $\gamma = 0.11$ ($\chi^2 = 0.81$). The constant-$H$ cuts (g--i) and constant-$K$ cuts (j--l) show direct comparison of fits with the data. The filled circles with error bars representing one standard deviation are the measured intensity and the blue solid lines are the fit. }
\label{supp10}
\end{figure}

\begin{figure}[h]
\includegraphics[width=0.75\textwidth]{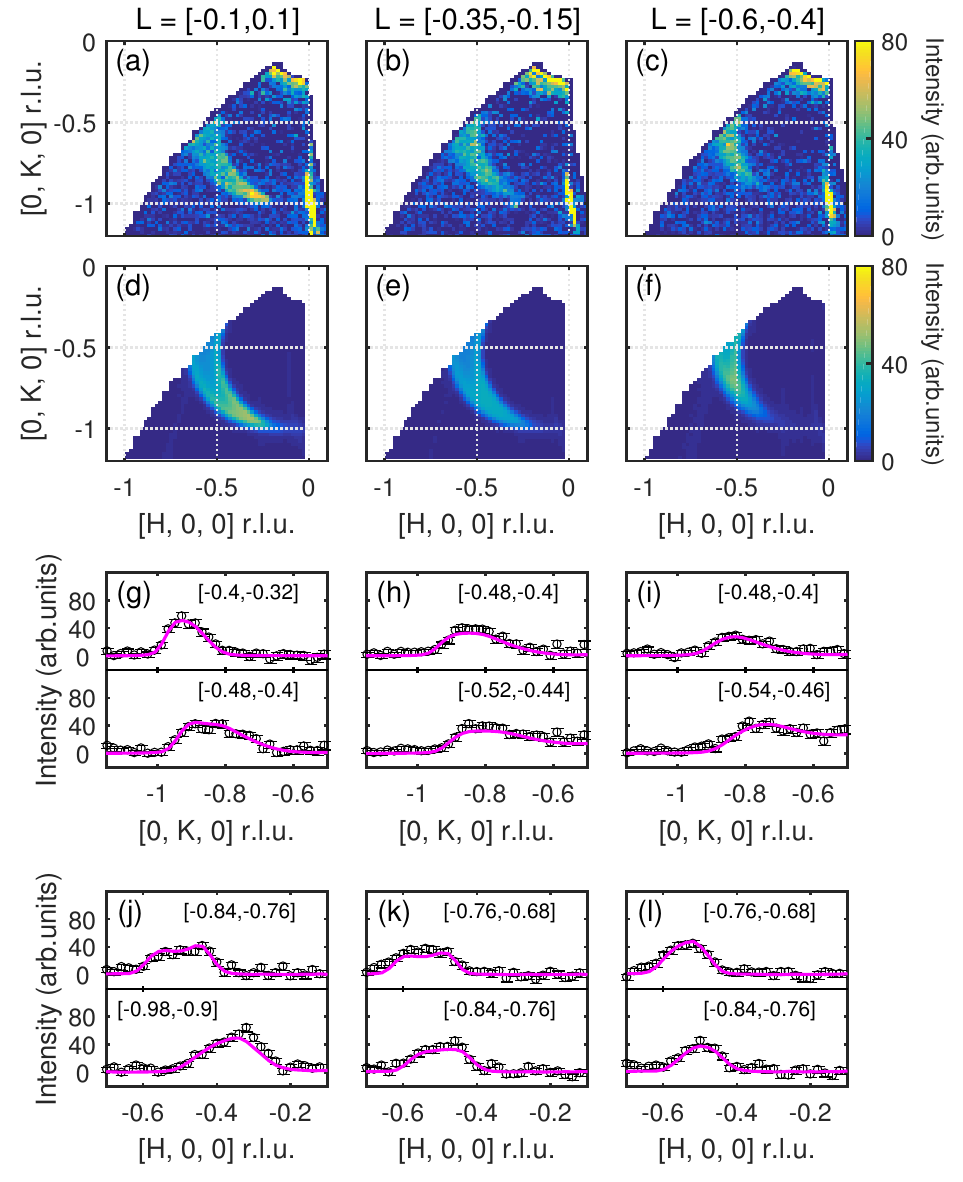}
\caption{Projected spin wave spectra of \cms\ measured by diffraction at $T = 50$~K. The data was averaged in the range $L = [-0.1,0.1]$ (a), $[-0.35,-0.15]$ (b), and $[-0.6,-0.4]$ (c). The direct beam plus constant background intensity was evaluated by fitting to a 2D Gaussian profile and subtracted from the data. (d--f) are the corresponding fitted spectra with the optimized $\gamma = 1.7(2)$~meV ($\chi^2 = 0.71$). The constant-$H$ cuts (g--i) and constant-$K$ cuts (j--l) show direct comparison of fits with the data. The filled circles with error bars representing one standard deviation are the measured intensity and the magenta solid lines are the fit. }
\label{supp11}
\end{figure}

\end{widetext}

\end{document}